\documentclass[epj]{webofc}
\usepackage[colorlinks,linkcolor=black,urlcolor=blue,citecolor=blue]{hyperref}
\usepackage[varg]{txfonts}   
\usepackage{amsmath}

\usepackage{ifthen} 

\usepackage{lineno}  
\usepackage{xspace} 

\usepackage{graphicx}  
\graphicspath{{./figs/}}   

\usepackage{color}
\usepackage{colortbl}

\usepackage{amsmath} 
\usepackage{upgreek} 

\usepackage{booktabs}

\newboolean{uprightparticles}
\setboolean{uprightparticles}{false} 

\usepackage{xspace} 
\usepackage{upgreek}

\def\MagUp {\mbox{\em Mag\kern -0.05em Up}\xspace}

\ifthenelse{\boolean{uprightparticles}}%
{

 \def\Pmu         {\ensuremath{\upmu}\xspace}

 \def\Ppsi        {\ensuremath{\uppsi}\xspace}

 \def\PDelta      {\ensuremath{\Delta}\xspace}                 
 \def\PXi      {\ensuremath{\Xi}\xspace}                 
 \def\PLambda      {\ensuremath{\Lambda}\xspace}                 
 \def\PSigma      {\ensuremath{\Sigma}\xspace}                 
 \def\POmega      {\ensuremath{\Omega}\xspace}                 
 \def\PUpsilon      {\ensuremath{\Upsilon}\xspace}                 
 
 \def\PB      {\ensuremath{\mathrm{B}}\xspace}                 
                  
 \def\PD      {\ensuremath{\mathrm{D}}\xspace}

 \def\PJ      {\ensuremath{\mathrm{J}}\xspace}                 
 \def\PK      {\ensuremath{\mathrm{K}}\xspace}

 \def\Pb      {\ensuremath{\mathrm{b}}\xspace}

 \def\Pe      {\ensuremath{\mathrm{e}}\xspace}

 \def\Pi      {\ensuremath{\mathrm{i}}\xspace}

 \def\Ps      {\ensuremath{\mathrm{s}}\xspace}

}
{

 \def\Pmu         {\ensuremath{\mu}\xspace}

 \def\Ppsi        {\ensuremath{\psi}\xspace}                 
                  
 \mathchardef\PDelta="7101
 \mathchardef\PXi="7104
 \mathchardef\PLambda="7103
 \mathchardef\PSigma="7106
 \mathchardef\POmega="710A
 \mathchardef\PUpsilon="7107
                  
 \def\PB      {\ensuremath{B}\xspace}                 
                  
 \def\PD      {\ensuremath{D}\xspace}

 \def\PJ      {\ensuremath{J}\xspace}                 
 \def\PK      {\ensuremath{K}\xspace}

 \def\Pb      {\ensuremath{b}\xspace}

 \def\Pe      {\ensuremath{e}\xspace}

 \def\Pi      {\ensuremath{i}\xspace}

 \def\Ps      {\ensuremath{s}\xspace}

}

\makeatletter
\ifcase \@ptsize \relax
  \newcommand{\miniscule}{\@setfontsize\miniscule{4}{5}}
\or
  \newcommand{\miniscule}{\@setfontsize\miniscule{5}{6}}
\or
  \newcommand{\miniscule}{\@setfontsize\miniscule{5}{6}}
\fi
\makeatother

\DeclareRobustCommand{\optbar}[1]{\shortstack{{\miniscule (\rule[.5ex]{1.25em}{.18mm})}
  \\ [-.7ex] $#1$}}

\def\en         {{\ensuremath{\Pe^-}}\xspace}   
\def\ep         {{\ensuremath{\Pe^+}}\xspace}

\def\mup        {{\ensuremath{\Pmu^+}}\xspace}
\def\mun        {{\ensuremath{\Pmu^-}}\xspace} 
\def\mumu       {{\ensuremath{\Pmu^+\Pmu^-}}\xspace}

\def\squark    {{\ensuremath{\Ps}}\xspace}

\def\bquark    {{\ensuremath{\Pb}}\xspace}
\def\bquarkbar {{\ensuremath{\overline \bquark}}\xspace}
\def\bbbar     {{\ensuremath{\bquark\bquarkbar}}\xspace}

\def\kaon    {{\ensuremath{\PK}}\xspace}
  \def\Kbar    {{\kern 0.2em\overline{\kern -0.2em \PK}{}}\xspace}

\def\KorKbar    {\kern 0.18em\optbar{\kern -0.18em K}{}\xspace}

\def\Kp      {{\ensuremath{\kaon^+}}\xspace}

\def\Kstar   {{\ensuremath{\kaon^*}}\xspace}

  \def\Dbar    {{\kern 0.2em\overline{\kern -0.2em \PD}{}}\xspace}

\def\DorDbar    {\kern 0.18em\optbar{\kern -0.18em D}{}\xspace}

\def\B       {{\ensuremath{\PB}}\xspace}
\def\Bbar    {{\ensuremath{\kern 0.18em\overline{\kern -0.18em \PB}{}}}\xspace}

\def\BorBbar    {\kern 0.18em\optbar{\kern -0.18em B}{}\xspace}

\def\Bu      {{\ensuremath{\B^+}}\xspace}

\def\Bp      {{\ensuremath{\Bu}}\xspace}

\def\Bd      {{\ensuremath{\B^0}}\xspace}
\def\Bs      {{\ensuremath{\B^0_\squark}}\xspace}

\def\jpsi     {{\ensuremath{{\PJ\mskip -3mu/\mskip -2mu\Ppsi\mskip 2mu}}}\xspace}

  \def\Y#1S{\ensuremath{\PUpsilon{(#1S)}}\xspace}

\def\Lbar        {{\ensuremath{\kern 0.1em\overline{\kern -0.1em\PLambda}}}\xspace}
\def\LorLbar    {\kern 0.18em\optbar{\kern -0.18em \PLambda}{}\xspace}

\def\BF         {{\ensuremath{\mathcal{B}}}\xspace}

\newcommand{\decay}[2]{\ensuremath{#1\!\to #2}\xspace}         

\def\to                 {\ensuremath{\rightarrow}\xspace}

\def\qsq       {{\ensuremath{q^2}}\xspace}

\def\AT#1     {\ensuremath{A_{\mathrm{T}}^{#1}}\xspace}           

\def\C#1      {\ensuremath{\mathcal{C}_{#1}}\xspace}                       
\def\Cp#1     {\ensuremath{\mathcal{C}_{#1}^{'}}\xspace}                    
\def\Ceff#1   {\ensuremath{\mathcal{C}_{#1}^{\mathrm{(eff)}}}\xspace}        
\def\Cpeff#1  {\ensuremath{\mathcal{C}_{#1}^{'\mathrm{(eff)}}}\xspace}       
\def\Ope#1    {\ensuremath{\mathcal{O}_{#1}}\xspace}                       
\def\Opep#1   {\ensuremath{\mathcal{O}_{#1}^{'}}\xspace}                    

\newcommand{\tev}{\ifthenelse{\boolean{inbibliography}}{\ensuremath{~T\kern -0.05em eV}\xspace}{\ensuremath{\mathrm{\,Te\kern -0.1em V}}}\xspace}
\newcommand{\gev}{\ensuremath{\mathrm{\,Ge\kern -0.1em V}}\xspace}
\newcommand{\mev}{\ensuremath{\mathrm{\,Me\kern -0.1em V}}\xspace}
\newcommand{\kev}{\ensuremath{\mathrm{\,ke\kern -0.1em V}}\xspace}
\newcommand{\ev}{\ensuremath{\mathrm{\,e\kern -0.1em V}}\xspace}
\newcommand{\gevc}{\ensuremath{{\mathrm{\,Ge\kern -0.1em V\!/}c}}\xspace}
\newcommand{\mevc}{\ensuremath{{\mathrm{\,Me\kern -0.1em V\!/}c}}\xspace}
\newcommand{\gevcc}{\ensuremath{{\mathrm{\,Ge\kern -0.1em V\!/}c^2}}\xspace}
\newcommand{\gevgevcccc}{\ensuremath{{\mathrm{\,Ge\kern -0.1em V^2\!/}c^4}}\xspace}
\newcommand{\mevcc}{\ensuremath{{\mathrm{\,Me\kern -0.1em V\!/}c^2}}\xspace}

\def\gsim{{~\raise.15em\hbox{$>$}\kern-.85em
          \lower.35em\hbox{$\sim$}~}\xspace}
\def\lsim{{~\raise.15em\hbox{$<$}\kern-.85em
          \lower.35em\hbox{$\sim$}~}\xspace}

\def\tell1  {TELL1\xspace}
\def\ukl1   {UKL1\xspace}

\newcommand{\matel}[3]{\langle #1|#2|#3\rangle}
\newcommand{\al}{\alpha}

\newcommand{\de}{\delta}

\newcommand{\V}{{\cal V}}

\wocname{EPJ Web of Conferences}

\woctitle{CONF12}

\begin{document}

\begin{flushright}
\begin{tabular}{l}
\small{CP3-Origins-2017-009 DNRF90} \\ 
\small{ICCUB-17-008} 
 \end{tabular}
\end{flushright}

\selectlanguage{english}
\title{Flavour  Anomalies in \boldmath{$b\to s\ell^+\ell^-$} Processes - \\
 a round table Discussion}

\author{
T.~Blake\inst{1}
\and M.~Gersabeck\inst{2}\fnsep\thanks{\email{Marco.Gersabeck@cern.ch}}
\and L.~Hofer\inst{3}
\and S.~J\"ager\inst{4}
\and Z.~Liu\inst{5}
\and R.~Zwicky\inst{6}
 }

\institute{
University of Warwick, Coventry, UK
\and The University of Manchester, Manchester, UK
\and Institut de Ciencies del Cosmos, Universitat de Barcelona, Barcelona, Spain
\and University of Sussex, Brighton, UK 
\and Institute of High Energy Physics and Theoretical Physics Center for Science Facilities, Chinese Academy of Sciences, Beijing, 100049, China
\and Higgs Centre for Theoretical Physics, University of Edinburgh, Edinburgh, UK
}

\abstract{
  Precision measurements of flavour observables  provide powerful tests of many extensions of the Standard Model. 
  This contribution covers a range of flavour measurements of $b\to s\ell^+\ell^-$ transitions, several of which are in tension with the Standard Model of particle physics, as well as their theoretical interpretation.
  The basics of the theoretical background are discussed before turning to the main question of the field: whether the anomalies can be explained by QCD effects or whether they may be indicators of effects beyond the Standard Model.
}

\maketitle

\newpage
\setcounter{tocdepth}{1}
\tableofcontents

\section{Introduction} 
\label{sec:introduction}

Flavour physics has a long track record of discoveries that paved the way for advances in particle physics.
In particular the discovery of the \Bd meson oscillations in 1987~\cite{Prentice:1987ap} is a great example  demonstrating the potential of flavour physics to infer physics of high mass scales through precision measurements at low scales: the observed rate of oscillations was the first indication of the top quark 
being much heavier than the other five quark flavours.
Precision flavour measurements at the LHC are sensitive to indirect effects from physics beyond the Standard Model (SM) at far greater scales than those accessible in direct searches.

This paper summarises a panel discussion focussing on a range of anomalies seen in measurements of 
\decay{\bquark}{\squark \ell^+ \ell^-} decays, which are a sensitive class of processes to potential effects 
beyond the SM (BSM).
The central question is whether the observed anomalies 
 are indeed BSM effects or 
 whether they can be explained by QCD effects. The general \decay{\bquark}{\squark \ell^+ \ell^-} 
 framework is briefly discussed below, 
  the experimental situation is reviewed in Sec.~\ref{sec:experiment},  
  which are 
  followed by theory   
Secs.~\ref{sec:observables}-\ref{sec:charm} on the heavy quark framework, 
form-factor determinations and the relevance of long distance charm contributions.  
Global fit strategies aiming to combine experimental observables to extract a more precise theoretical picture 
are presented in Sec.~\ref{sec:global},  and finally 
 Sec.~\ref{sec:bsm} covers potential BSM interpretations of the flavour anomalies before the summary, which is given in Sec.~\ref{sec:summary}.

\subsection{General $\decay{\bquark}{\squark \ell^+ \ell^-}$ framework}
\label{sec:general}

In order to maintain some coherence between the different contributions of various authors 
we give the effective Hamiltonian 
with some minimal explanation referencing to the theory based sections. 
\begin{equation}
\label{eq:Heff}
\mathcal H_{\mathrm{eff}} = \frac{G_F}{\sqrt 2}\left( \sum_{i=1}^2 
(\lambda_u C_i O_i^u + \lambda_c C_i \mathcal  O_i^c )       -\lambda_t \sum_{i=3}^{10} C_i  O_i \right)  \;, \qquad 
\lambda_i \equiv V_{\text{is}}^*V_{\text{ib}} \;,
\end{equation}
where $V_{\text{ij}}$ are CKM-elements, $\mu$ is a factorisation scale,  $C_i$ the Wilson coefficients encoding the ultraviolet (UV) physics and some of the most important  operators are 
\begin{alignat}{4}
\label{eq:SMbasis}
& \mathcal O_1^q &\;=\;&  (\bar s_i q_j)_{V-A}(\bar q_j b_i)_{V-A} \;,  \qquad  \qquad 
& & \mathcal O_2^q &\;=\;&  (\bar s_i q_i)_{V-A}(\bar q_j b_j)_{V-A}  \;, \nonumber \\
& \mathcal O_{7[8]} &\;=\;& -\frac{e m_b}{8\pi^2}\bar s \sigma\cdot F[G] (1+\gamma_5)b  \;,
& & \mathcal O_{9,10} &\;=\;& \frac{\alpha}{2\pi}(\bar \ell\gamma^\mu [\gamma_5] \ell)(\bar s \gamma_\mu (1-\gamma_5) b)  \;, 
\end{alignat}
where $i,j$ are colour indices, $(\bar s b)_{V\pm A}=\bar s \gamma^\mu (1\pm\gamma_5) b$,  $e = \sqrt{4 \pi \alpha} > 0 $  and $G_F$ is the Fermi constant.  The Wilson coefficients $C_i$ are computed to 
next-to-next-to-leading order (NNLO) in perturbation theory, and the matrix elements of the operators 
are of non-perturbative nature. Those are either local short distance form-factors or non-local long distance 
matrix elements.  
The form-factors are computed in LCSR
and in lattice QCD in complementary regimes as discussed in 
Sec.~\ref{sec:formfactors}, and obey some 
symmetry relations for large  large $m_b$-mass, cf. Secs~\ref{sec:observables} and \ref{sec:eom} 
respectively. 
At large recoil of the lepton pair  
the matrix elements are evaluated outside the resonance region,  
either within the heavy quark expansion  or LCSR
as discussed in Secs.~\ref{sec:observables} and \ref{sec:charm} respectively. 
At low-recoil an OPE has been proposed  
cf. Sec.~\ref{sec:observables}, subject to potentially significant corrections from 
resonant charm contribution as discussed  in Sec.~\ref{sec:charm}.

\section{Summary of experimental situation (T. Blake)}
\label{sec:experiment}

There has been huge experimental progress in measurements of rare $b\to s$ processes in the past five years. 
This has been driven by the large \bbbar production cross-section in $pp$ collisions at the LHC, 
which enabled the LHC experiments to collect unprecedented samples of decays with dimuon final-states. 

\subsection{Leptonic decays} 

The decay \decay{\Bs}{\mumu} is considered a golden mode for testing the SM at the LHC. 
The SM branching fraction depends on a single hadronic parameter, the \Bs decay constant, that be computed from Lattice CQD. 
Consequently the SM branching fraction is known to better than a 10\% precision~\cite{Bobeth:2013uxa}.
A combined analysis of the CMS and LHCb datasets~\cite{LHCb-PAPER-2014-049} results in a time-averaged measurement of the branching fraction of the decay of
\begin{align}
\overline{\BF}(\decay{\Bs}{\mumu}) = (2.8^{+0.7}_{-0.6})\times 10^{-9}~.
\end{align}
A recent ATLAS measurement~\cite{Aaboud:2016ire} yields
\begin{align}
\overline{\BF}(\decay{\Bs}{\mumu}) = (0.9^{+1.1}_{-0.8})\times 10^{-9}~, 
\end{align}
which is consistent with the combined analysis from CMS and LHCb.
These measurements are in good agreement with SM predictions and set strong constraints on extensions of the SM that introduce new scalar or pseudoscalar couplings.

\subsection{Semileptonic decays}

\begin{figure}
\centering
\includegraphics[width=0.6\linewidth]{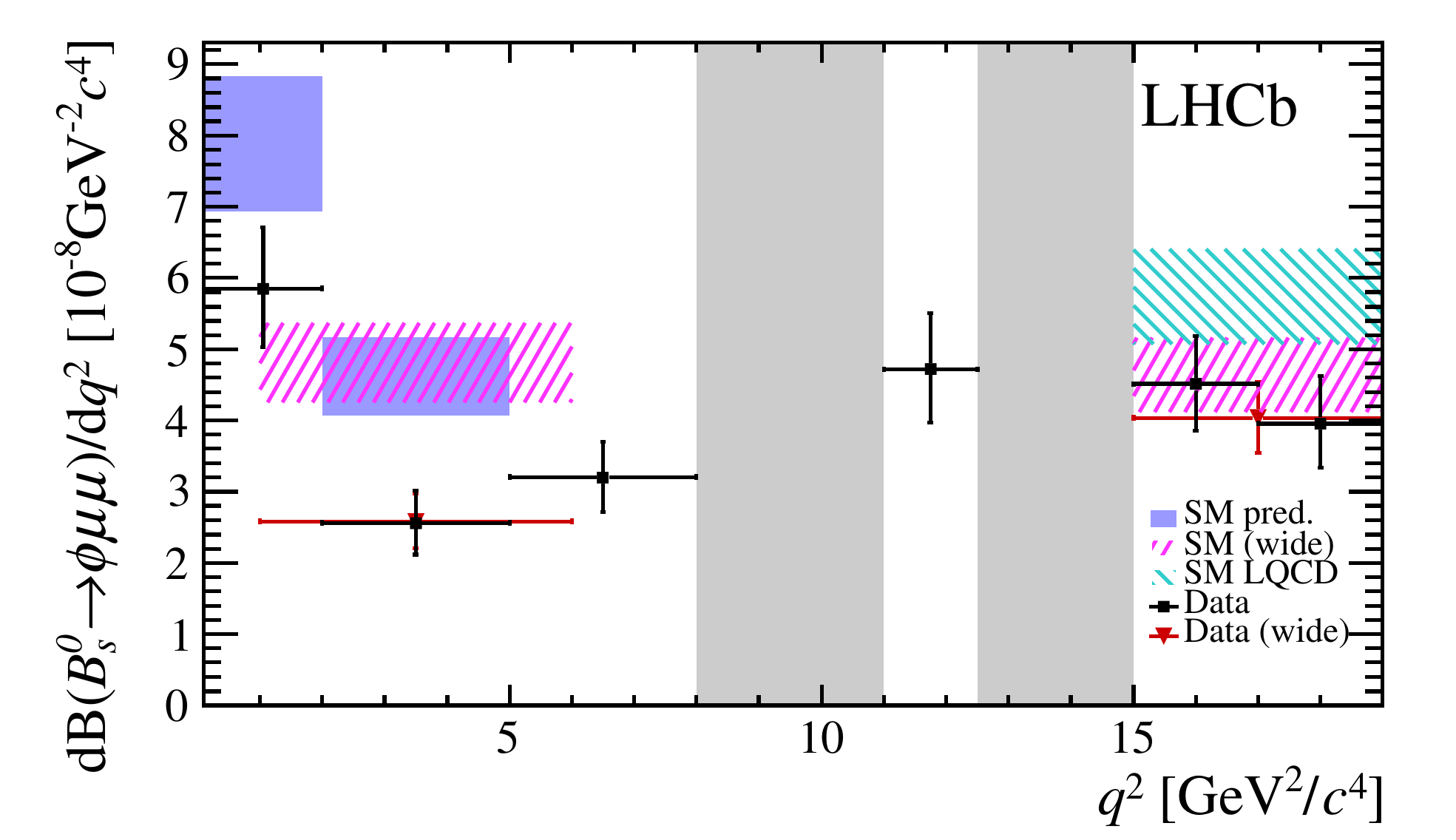}
\caption{ 
Differential branching fraction of the \decay{\Bs}{\phi\mumu} decay measured by the LHCb experiment~\cite{LHCb-PAPER-2015-023} as a function of the dimuon invariant mass squared, \qsq. 
The data are compared to SM predictions based on Refs.~\cite{Altmannshofer:2014rta,Straub:2015ica} and \cite{Horgan:2013pva}.
The rise in the branching fraction at low \qsq arises from virtual photon contributions to the decay. Reproduced from Ref.~\cite{LHCb-PAPER-2015-023}.
\label{fig:branching}
}
\end{figure}

The large dataset has also enabled the LHCb and CMS experiments to make the most precise measurements of the differential branching fraction of  \decay{\B}{K^{(*)}\mumu} and \decay{\Bs}{\phi\mumu} to date~\cite{LHCb-PAPER-2014-006, LHCb-PAPER-2015-023, Khachatryan:2015isa}. 
Above the open charm threshold, broad resonances are seen in the LHCb dataset~\cite{LHCb-PAPER-2013-039}.  
The most prominent of these is the $\psi(4160)$. 
The regions close to the narrow charmonium resonances are excluded from the analysis. 
With the present binning scheme, the uncertainties on differential branching fraction measurements are limited by the knowledge of the \decay{\B}{\jpsi K^{(*)}} and \decay{\Bs}{\jpsi\phi} branching fractions that are used to normalise the signal. 
The measured differential branching fractions of $b \to s \mumu$ processes, tend to prefer smaller values than their corresponding SM predictions. 
The largest discrepancy is seen in the \decay{\Bs}{\phi\mumu} decay, where the data are more than $3\,\sigma$ from the SM predictions in the dimuon invariant mass squared range $1 < \qsq < 6\gev^{2}/c^4$, see Fig~\ref{fig:branching}.

While the branching fractions of rare \decay{B}{K^{(*)}\mumu} decays have large theoretical uncertainties arising from the $\B \to K^{(*)}$ form-factors, many sources of uncertainty will cancel when comparing the decay rates of the \decay{B}{K^{(*)}\mumu} and \decay{B}{K^{(*)}\ep\en} decays. 
In the range $1 < q^2 < 6\gev^2/c^4$, the LHCb experiment measures~\cite{LHCb-PAPER-2014-024}
\begin{align}
\label{eq:RK}
R_{K}[1,6] = 0.745 \,^{+0.090}_{-0.074} ({\rm stat}) \,^{+0.035}_{-0.035}({\rm syst})\,.
\end{align} 
This is approximately $2.6\,\sigma$ from the SM expectation of almost identical decay rates for the two channels. 
In order to cancel systematic differences between the reconstruction of electrons and muons in the detector, the LHCb analysis is performed as a double ratio to the rate  \decay{\Bp}{\jpsi\Kp} decays (where the \jpsi can decay to a dielectron or dimuon pair).  
The migration of events in $q^2$ due to final-state-radiation is accounted for using samples of simulated events. QED effects are simulated through PHOTOS~\cite{Golonka:2005pn}. 
The largest difference between the dimuon and dielectron final-states comes from Bremsstrahlung from the electrons in the detector. 
This is simulated using GEANT\,4~\cite{Agostinelli:2002hh}.  
The modelling of the migration of events and the line-shape of the decay are the main contributions to the systematic uncertainty on the measurement. 

\begin{figure}
\centering
\includegraphics[width=0.48\linewidth]{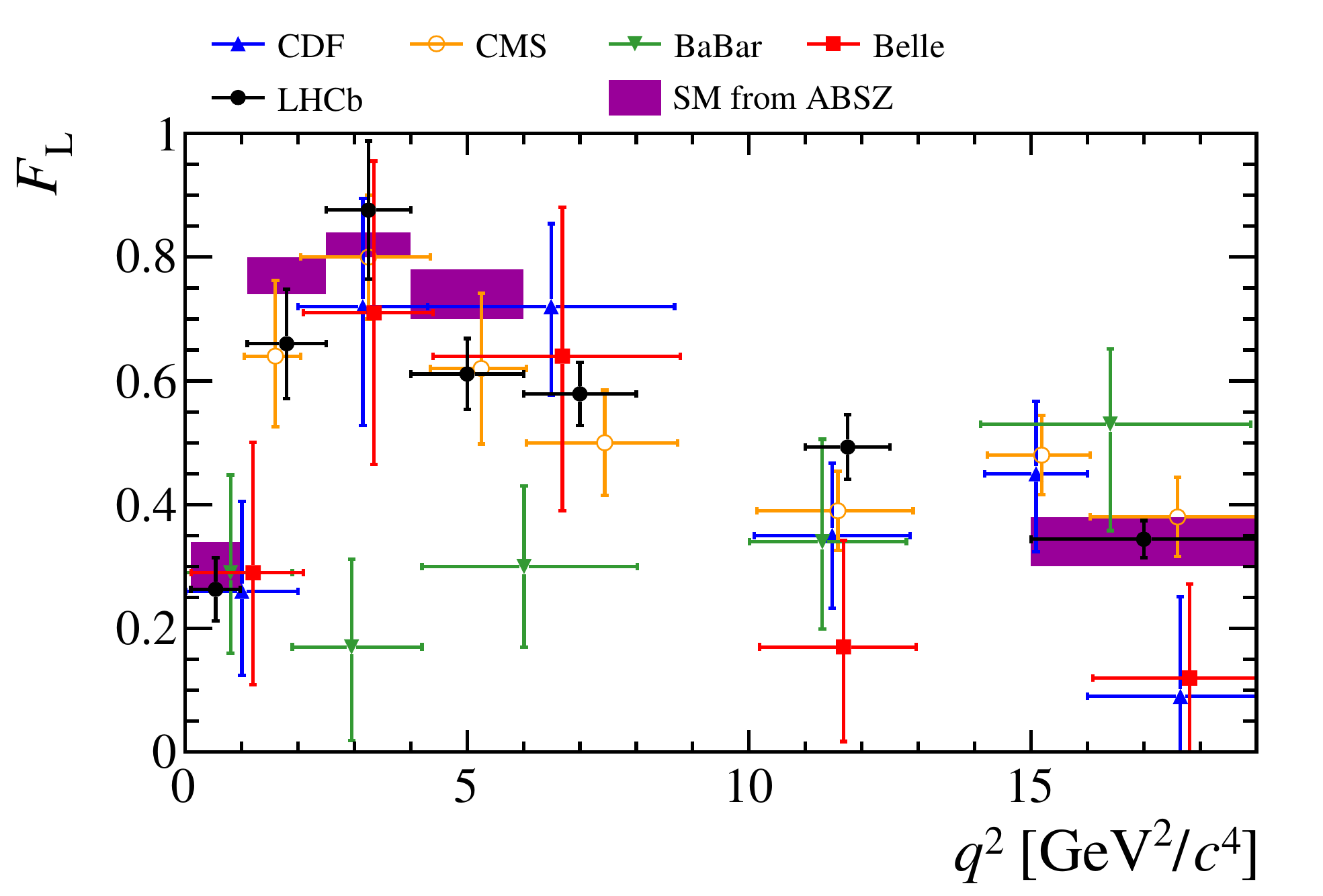}
\includegraphics[width=0.48\linewidth]{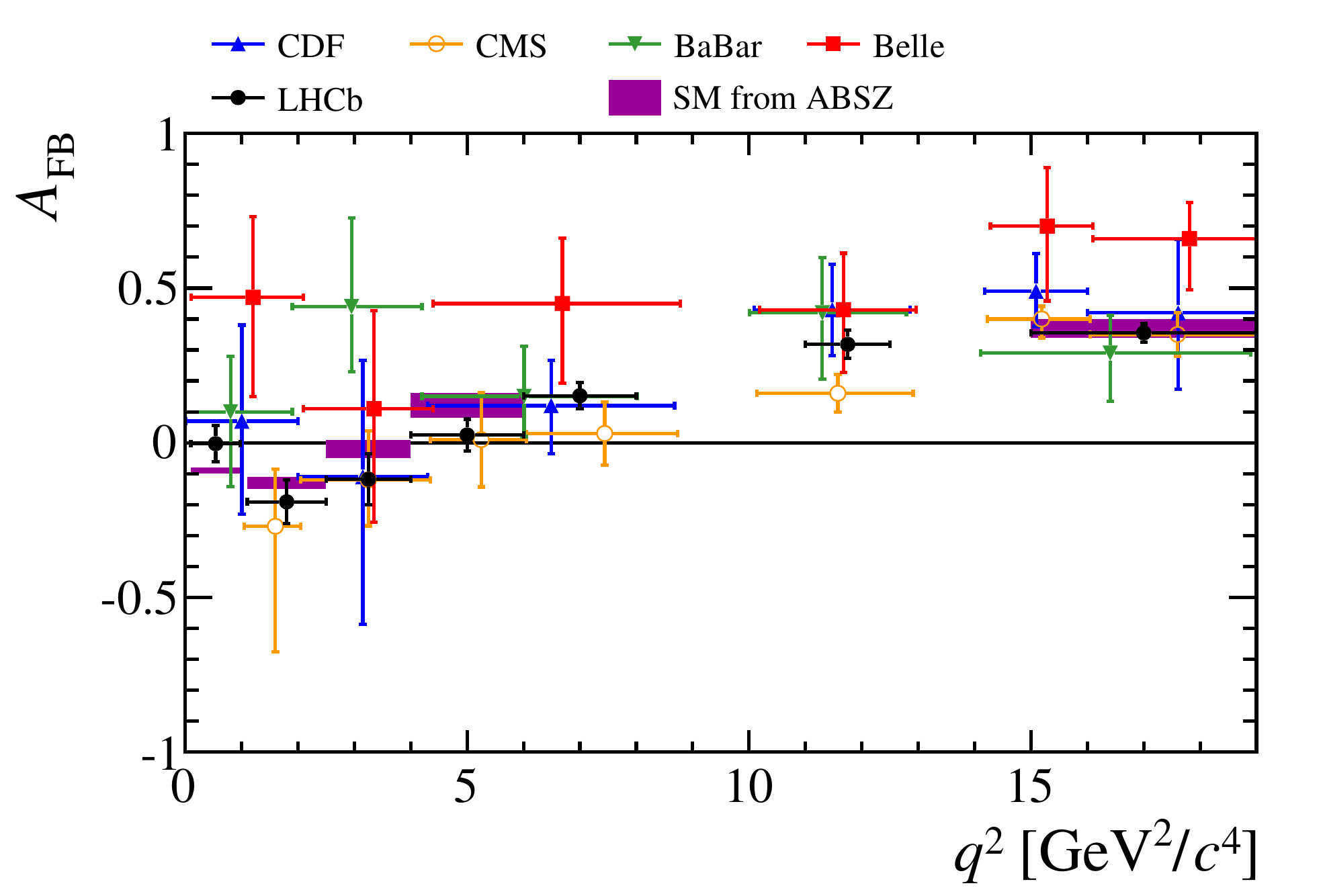}
\caption{
Observables $F_{\rm L}$ and $A_{\rm FB}$ measured by the BaBar~\cite{Lees:2015ymt}, Belle~\cite{Wei:2009zv}, CDF~\cite{Aaltonen:2011ja}, CMS~\cite{Khachatryan:2015isa} and LHCb~\cite{LHCb-PAPER-2015-051} experiments for the \decay{B}{\Kstar\mumu} decay as a function of the dimuon invariant mass squared, \qsq. The shaded region indicates a theoretical prediction for the observables based on Refs.~\cite{Altmannshofer:2014rta,Straub:2015ica}.
No data point is shown for CMS in the range $q^2 < 1\gev^{2}/c^{4}$, due to the thresholds used in the CMS trigger system.
\label{fig:kstarmumu:obs}
}
\end{figure}

The distribution of the final-state particles in the \decay{B}{\Kstar\ell^+ \ell^-} decay can be described by three angles and $q^2$. 
The angles are: the angle between the direction of the $\ell^+$ ($\ell^-$) and the \B (\Bbar) in the rest-frame of the dilepton pair; the angle between the direction of the kaon and the direction of the \B in the \Kstar rest-frame; and the angle between the decay planes of the dilepton pair and the \Kstar in the rest-frame of the \B, denoted $\phi$. 
The resulting angular distribution can be parameterised in terms of eight angular observables: the longitudinal polarisation of the \Kstar, $F_{\rm L}$; the forward-backward asymmetry of the dilepton system, $A_{\rm FB}$; and six additional observables that cancel when integrating over $\phi$. 
Existing measurements of the observables $A_{\rm FB}$ and $F_{\rm L}$ are shown in Fig.~\ref{fig:kstarmumu:obs} along with SM predictions based on Refs.~\cite{Altmannshofer:2014rta,Straub:2015ica}. 
The most precise measurements of the $F_{\rm L}$ and $A_{\rm FB}$ come from the LHCb and CMS experiments.
In general, the measurements are consistent with each other and are compatible with the SM predictions. 
The largest tension is seen in the BaBar measurement of $F_{\rm L}$~\cite{Lees:2015ymt}.

\begin{figure}
\centering
\includegraphics[width=0.6\linewidth]{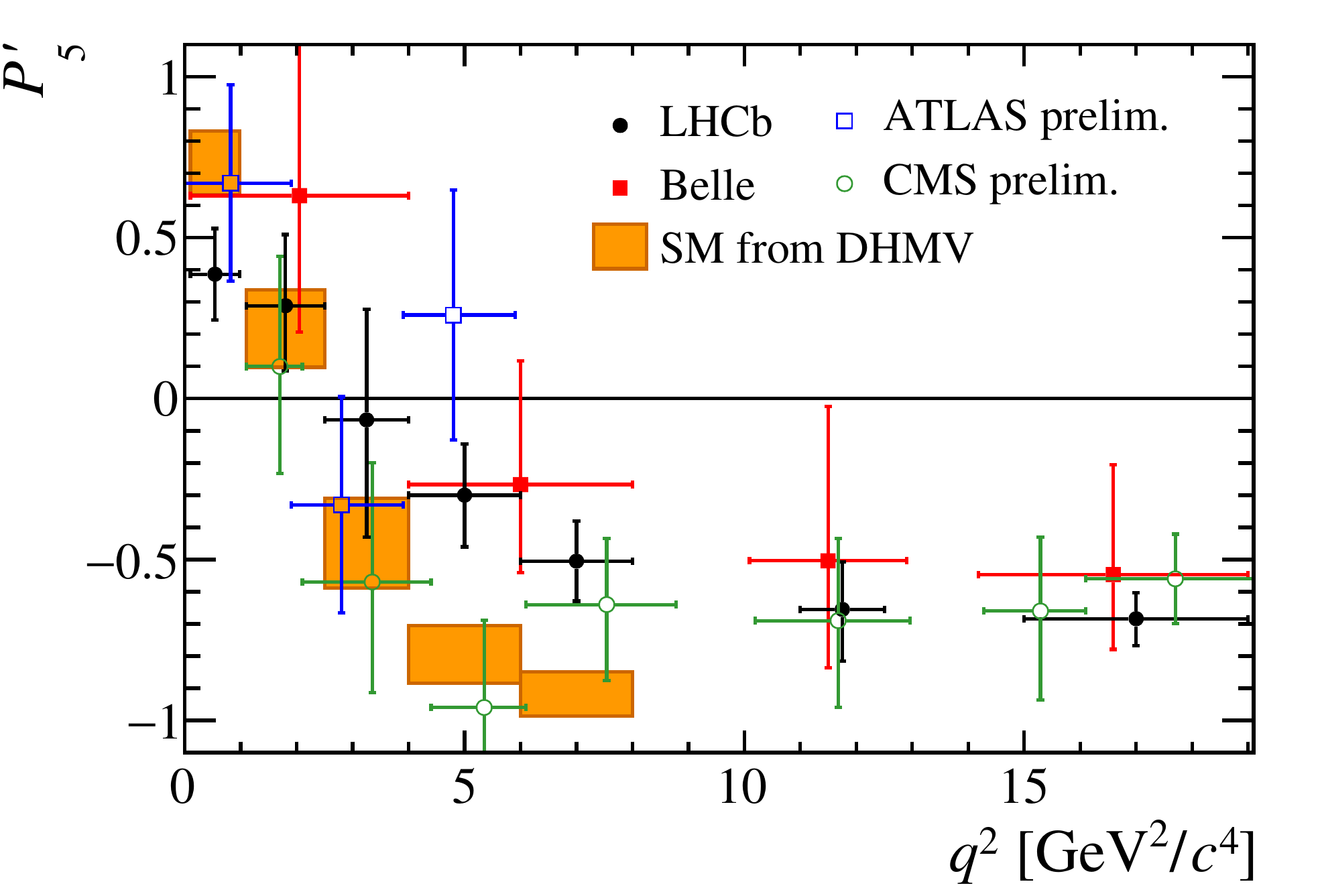}
\caption{
Observable $P_5'$ measured by LHCb~\cite{LHCb-PAPER-2015-051} and Belle~\cite{Abdesselam:2016llu} as a function of the dimuon invariant mass squared, \qsq, in the \decay{B}{\Kstar\mumu} decay. Preliminary results from ATLAS~\cite{ATLAS-CONF-2017-023} and CMS~\cite{CMS-PAS-BPH-15-008} are also included. The shaded regions indicate theoretical predictions from Ref.~\cite{Descotes-Genon:2014uoa}. 
\label{fig:kstarmumu:p5p}
}
\end{figure}

The ATLAS, CMS, LHCb and Belle experiments have also measured the remaining angular observables that are usually cancelled by integrating over $\phi$. 
The LHCb collaboration  performed a first full angular analysis of the decay in Ref.~\cite{LHCb-PAPER-2015-051}.
The majority of these additional observables are consistent with SM predictions.  
However, a tension exists between measurements of the observable $P_5'$ and their corresponding SM prediction in the region $4 < q^2 < 8\gev^2/c^4$. 
This tension is illustrated in Fig.~\ref{fig:kstarmumu:p5p}. 
In the region $4 < \qsq < 8\gev^{2}/c^{4}$, the data from ATLAS, Belle and LHCb are significantly above the SM predictions. 
The CMS result is more consistent. 

The experimental measurements of the angular observables are currently statistically limited. 
The largest sources of systematic uncertainty arise from modelling of the experimental angular acceptance and the background angular distribution. 

\section{Rare $B$-decays and the heavy quark expansion (S. J\"ager)}
\label{sec:observables}

\subsection{Context}
In the Standard Model (SM), the amplitude for a rare semileptonic
decay $\bar B \to M \ell^+ \ell^-$,
with $M$ a hadronic system such as $\bar K$, $\bar K \pi, \dots$
can be written, to leading order in the electromagnetic
coupling $\alpha_{\rm EM}$ but exact in QCD, as
\begin{equation}
  {\cal A}(B \to M \ell^+ \ell^-) =
    L^\mu a_{V\mu} + L^{5\mu} a_{A\mu} ,
\end{equation}
where $L^\mu$ and $L^{5\mu}$ are the vector and axial lepton currents.
One can trade $a_{V\mu}$, $a_{A\mu}$ for helicity amplitudes
$H_{V\lambda}(q^2)$
and $H_{A\lambda}(q^2)$,
where $\lambda$ is the helicity of the dilepton, which coincides with
  that of the hadronic state. For a kaon, $\lambda=0$ and there are two hadronic
  amplitudes. For a narrow $K^*$, $\lambda = \pm 1$ and there are six
  amplitudes.\footnote{We neglect the lepton mass,
    as the anomalies occur
    at $q^2 \gg m_\ell^2$. This eliminates a seventh amplitude. We will
    also neglect the strange quark mass and CKM-suppressed terms throughout.}
  These amplitudes determine all rate and angular observables measured
  at $B$-factories and LHCb (where the lepton spins are not measured).

In the conventions of \cite{Jager:2012uw},
$  H_{A\lambda} \propto C_{10} V_\lambda(q^2), $
where $V_\lambda$ are form factors and $C_{10}$ is the axial semileptonic
Wilson coefficient. (We omitted a normalisation
free of hadronic uncertainties.) Note that  $C_{10}$ and $V_\lambda$ are
renormalisation-scale and -scheme-independent.
$C_{10}$, in fact, receives
no contributions from below the weak scale at all
--- this is what makes
$B_s \to \mu^+ \mu^-$ such a precision observable.
The vector amplitudes can be written \cite{Jager:2012uw}
\begin{align} \label{eq:HVpm}
  H_{V\pm} &\propto \left[ C_9(\mu) V_{\pm}(q^2)
    + \frac{2m_b m_B}{q^2} C_7^{\rm eff}(\mu) T_{\pm}(q^2, \mu)
    - 16 \pi^2 \frac{m_B^2}{q^2} h_\pm(q^2, \mu) \right], \\
  H_{V0} &\propto \frac{\lambda^{1/2}}{2 m_B \sqrt{q^2}}
    \left[ C_9(\mu) V_0(q^2)
    + \frac{2 m_b}{m_B} C_7^{\rm eff}(\mu) T_0(q^2, \mu) \right]
     - 16 \pi^2 \frac{m_B^2}{q^2} h_0(q^2, \mu) .
\end{align}
Here $C_9(\mu)$ denotes the vector semileptonic Wilson coefficient,
$T_\lambda$ tensor form factors, multiplied by the ``effective'' dipole
Wilson coefficient $C_7^{\rm eff}$,
and 
$h_\lambda \sim \langle M(\lambda) |
T\left\{ j^{\rm em,had}(y) {\cal H}_{\rm had}^{\rm eff}(0) \right\} |
\bar B \rangle$ denotes the contribution from the hadronic
weak Hamiltonian (see  \cite{Jager:2012uw} for a precise definition).
So far everything is exact from a QCD point of view. The theoretical
difficulty resides in the evaluation of $h_\lambda$, which the heavy-quark
expansion achieves, and of the form factors, for which the heavy-quark
expansion provides relations.

One notes that
$C_9(\mu)$ and $C_7^{\rm eff}(\mu)$ are strongly scale-dependent.
  E.g., at NNLL order, $C_9(10 {\rm GeV}) = 3.75$, $C_9(5 {\rm GeV})=4.18$,
  $C_9(2.5 {\rm GeV}) = 4.49$.
  (One may compare this variation to a putative  $\Delta C_9^{\rm BSM} \sim -1$.)
  The scale dependence  must be precisely cancelled by
  the non-perturbative  object $h_\lambda$.
 (This follows from RG-invariance of the Hamiltonian, which is exact.)
  Any quantitative theoretical description of $h_\lambda$ must correctly
  incorporate this.\footnote{
  A recent LHCb paper \cite{Aaij:2016cbx} models $h_\lambda$, for $B \to K$,
  as a sum of Breit-Wigner
  resonances. While this more or less fits the data ($p \approx 0.01$),
  it has no scale-dependence, which is one reason why the coefficient
  ${\cal C}_9$ in that paper cannot be identified with $C_9(\mu)$ for any
  scale $\mu$.}

\subsection{The heavy-quark expansion}
\label{sec:heavy}

In the heavy-quark, large-recoil limit $E_M, m_B \gg \Lambda$, where
$\Lambda$ is the QCD scale parameter,
$h_\lambda$ factorizes into perturbative  hard-scattering kernels
multiplying form factors and light-cone distribution
amplitudes for the light meson $M = K, K^*$
\cite{Beneke:2001at,Bosch:2001gv,Beneke:2004dp}, known as QCD factorization.
It applies in the $q^2$-region below the narrow charm resonances, and perhaps
above the charm threshold up to $q^2 \sim 15 {\rm GeV}^2$.
QCDF can be formulated in soft-collinear effective field theory (SCET)
language \cite{Becher:2005fg}. ``Factorization'' is used in the Wilsonian
sense of separating physics of the scales $\sqrt{\Lambda m_b}$, $m_b$,
and above from the physics
of the scale $\Lambda$, not to be confused with ``naive factorization.''
There are two distinct classes
of effect. Vertex corrections may be compactly written as
\begin{eqnarray} \label{eq:C9eff}
  C_9 &\to& C_9(\mu) + Y(\mu, q^2, m_c) + \frac{\alpha_s}{4\pi} Y^{(1)}(\mu, q^2, m_c), \\ \label{eq:C7eff}
  C_7^{\rm eff} &\to& C_7^{\rm eff}(\mu) + \frac{\alpha_s}{4\pi} Z^{(1)}(\mu, q^2, m_c),
\end{eqnarray}
where $Y, Y^{(1)}$,and  $Z^{(1)}$ contain loop functions and Wilson coefficients.
This form makes the helicity independence of these effects evident. The
combination $C_9 + Y$ is traditionally called $C_9^{\rm eff}(q^2)$; the
heavy-quark limit justifies its use, but also predicts model-independent
higher-order corrections.
(Note that $C_9$ starts at ${\cal O}(1/\alpha_s)$ in the logarithmic counting,
so $Y$ is formally a NLL correction.) The
r.h.s.\ of (\ref{eq:C9eff}),(\ref{eq:C7eff}) are $\mu$-independent
up to ${\cal O}(\alpha_s^2)$ corrections.
The other class of effects constitutes so-called hard spectator scattering
and includes an annihilation contribution at ${\cal O}(\alpha_s^0)$, though
the latter comes with small CKM-factors and Wilson coefficients. Spectator
scattering probes
the structure of the $B$ and $K^*$-mesons through their light-cone distribution
amplitudes and is helicity dependent (and vanishes for $\lambda=+$).
Schematically,
\begin{equation} \label{eq:hspec}
  h_\lambda^{\rm spec} \propto T_\lambda(\alpha_s) * \phi_{B_\pm} * \phi_{K^*} .
\end{equation}
This expression is separately scale-independent, resulting in
formally $\mu$-independent observables.
Corrections to the heavy-quark limit scale like $\Lambda/m_B$ and
do not factorize, and must be estimated in other ways.

The expressions so far are sufficient to express all observables in terms
of form factors \cite{Altmannshofer:2008dz}, one can then
use form factor results from light-cone sum rules to compute observables.
Alternatively, one can make use of the fact that the
heavy-quark large-recoil limit also implies relations between
different form factors
\cite{Charles:1998dr,Beneke:2000wa,Beneke:2003pa}.
They look extremely simple in the helicity basis,
\begin{equation}   \label{eq:HQEFF}
T_\lambda(q^2) = V_\lambda(q^2) [1 + f_\lambda(q^2, \alpha_s)] +
   \mbox{spectator scattering} + {\cal O}(\Lambda/m_B)
\end{equation}
for $\lambda = -,0$ and $T_+(q^2) = V_+(q^2) = {\cal O}(\Lambda/m_B)$.
(The latter together with $h_+ = 0$ implies $H_{V+} = H_{A+} = 0$.)
Here $f_\lambda$ is a perturbative expression starting at ${\cal O}(\alpha_s)$.
The spectator-scattering contribution has a form similar to (\ref{eq:hspec}),
and is again proportional to $\alpha_s$. All ${\cal O}(\alpha_s^2)$ corrections
are also known \cite{Beneke:2004rc,Hill:2004if,Becher:2004kk,Beneke:2005gs}.
Note that the ratios $T_\lambda/V_\lambda$ are free of hadronic input
in the heavy-quark limit, up to (calculable) spectator-scattering
and (incalculable) power corrections.

There is an alternative expansion, applicable at $M_B^2 , q^2 \gg \Lambda^2$,
in particular above the $D \bar D$ threshold.
The actual expansion is in $E_M/\sqrt{q^2}$ and expresses $h_{\lambda}$ in terms
of matrix elements of local operators (OPE) of increasing dimension, again
with perturbatively calculable coefficients
\cite{Buchalla:1998mt,Grinstein:2004vb,Beylich:2011aq}.
This is not by itself a heavy-quark expansion, although the kinematics ensure
that a HQE is valid and the $b$-quark field may be expanded accordingly
\cite{Grinstein:2004vb}, and the HQE can be used to estimate the
matrix elements \cite{Beylich:2011aq}.
The leading matrix elements are again the form factors $T_\lambda$ and $V_\lambda$,
with perturbative coefficient functions that coincide with $C_7^{\rm eff}$
and $C_9^{\rm eff}(q^2)$ but different $\alpha_s$-corrections. The leading
higher-dimensional corrections are of order $\Lambda^2/q^2$ and negligible, in
particular spectator scattering is (strongly) power-suppressed.
The OPE also gives a qualitative picture how open-charm resonances arise,
by means of analytic continuation of OPE remainder terms from spacelike to
physical (timelike) $q$, 
$ \exp(-c \sqrt{-q^2}/\Lambda) \to \exp(-i c \sqrt{q^2}/\Lambda)$, giving
oscillatory behaviour. Note that these ``duality-violating'' terms
are nonanalytic in $\Lambda$,
hence the large-recoil $\Lambda/m_B$ expansion will not capture them either.
Formally such
terms are of ``infinite order'' in $\Lambda/\sqrt{q^2}$ or $\Lambda/m_B$.
They become less important away from the threshold and partly cancel out in
binned observables (see \cite{Beylich:2011aq,Lyon:2014hpa}).
See also the discussion of charm resonances
in $B \to K^{(*)} \mu^+ \mu^-$ in Section \ref{sec:charm}.

\subsection{Phenomenology}
How large are the various effects and residual uncertainties?
The shift to $C_7^{\rm eff}$ is an ${\cal O}(25\%)$ constructive correction.
The ${\cal O}(\alpha_s)$ correction to $C_9$ is an order $5\%$ destructive effect
which partly cancels the ${\cal O}(\alpha_s^0)$ term i.e.\ $Y(q^2)$.
The spectator-scattering corrections are smaller, though the normalisation
is quite uncertain (by about a factor of two) due to the poor knowledge of
the $B$-meson LCDA $\phi_{B_+}$. Overall the effects are significant.
For instance, $\lim_{q^2 \to 0} [q^2 H_{V-}(q^2) ]$ gives the $B \to K^* \gamma$
amplitude, which receives a $+30\%$ correction, or a $+70\%$ correction
at the rate level. Because $C_7^{\rm eff}$ is tightly
constrained from the measured $B \to X_s \gamma$ rate, this allowed
\cite{Beneke:2001at} to conclude that, absent large power corrections,
$T_-(0) = T_1(0) = 0.27 \pm 0.04 $, updated to $T_1(0) = 0.28 \pm 0.02$ in
\cite{Beneke:2004dp} and well below LCSR predictions at the time.
A recent LCSR evaluation \cite{Straub:2015ica}
gives $T_1(0) = 0.308 \pm 0.031$.
The consistency supports smallness of power corrections.

It is impossible to give a comprehensive phenomenology of angular
$B \to K^* \mu^+ \mu^-$ observables
in this space. Two observables that are very sensitive to $C_9$ are the
forward-backward asymmetry,
specifically its zero-crossing, and the angular term $S_5$ (or $P_5'$).
As long as Wilson coefficients are real, the FBAS zero is determined
by ${\rm Re}\, H_{V-}(q_0^2) = 0$, up to {\em second-order}
power corrections ${\cal O}(\Lambda^2/m_B^2)$. From (\ref{eq:HVpm})
it is clear that the zero depends on $(C_7^{\rm eff}/C_9) \times (T_-/V_-)$,
essentially free from form-factor uncertainties in the heavy-quark limit.
Given the fact that $C_7^{\rm eff}$ is essentially pinned to its SM value
by $B \to X_s \gamma$, a $q^2_0$-measurement may hence be viewed
as a determination of $C_9$. Ref.\ \cite{Beneke:2004dp}
obtained $q^2_0 = 4.36^{+0.33}_{-0.31}$ GeV${}^2$ (neglecting power corrections),
to be compared to the LHCb determination \cite{LHCb-PAPER-2015-051}
$q^2_0 \in [3.40, 4.87]$ GeV${}^2$, in good agreement.
By comparison, the LCSR form factors of \cite{Straub:2015ica}
imply a lower crossing point, around $3.5$ GeV${}^2$, giving a slight
preference for $C_9 < C_9^{\rm SM}$. This can be traced to a
ratio of $T_-/V_-$ that is ${\cal O}(10 \%)$ below the
heavy-quark-limit prediction, which is still consistent with a power correction.
The eventual relative accuracy on $C_9$ is limited
by that on ratio $T_-(q_0^2)/V_-(q_0^2)$.

The observable $P_5'$ \cite{Matias:2012xw}, which shows the most pronounced
anomaly, depends on all six helicity amplitudes. It hence depends on
ratios $T_-/V_-$, $T_0/V_0$, $V_+/V_-$, $T_+/V_-$, and $V_0/V_-$. The
last of these does not satisfy a heavy-quark relation, but $P_5'$ has been
constructed in such a way that dependence on it cancels in the heavy-quark
limit if $\alpha_s$-corrections are neglected.
It has been suggested to employ $V \propto V_+ + V_-$ instead
of $V_-$ \cite{Descotes-Genon:2014uoa}, which
reduces the explicit sensitivity to $V_+$
in $P_5'$. From a pure heavy-quark perspective,
the physics of $V_+$ (which involves soft physics flipping the helicity
of the strange quark emitted from $b$-quark decay, or changing the
helicity of the $B$-remnant absorbed into the $K^*$) and $V_-$ (which does
not require such a spin-flip) seem very different, such that the conservative
choice is to associate separate uncertainties to both (hence a larger one to
$V$). The impact on
the significance of the $P_5'$ anomaly is noticeable because of this
\cite{Jager:2014rwa,Descotes-Genon:2014uoa} and
because \cite{Descotes-Genon:2014uoa} adopt central values for
power corrections to match LCSR form factor central values while
\cite{Jager:2014rwa} use zero central values (the HQ limit).
Further recent investigations on the role of power corrections in
semileptonic $B$-decays can be found
in \cite{Ciuchini:2016weo,Capdevila:2017ert,Chobanova:2017ghn}.

\subsection{Discussion}
The heavy-quark expansion goes a long way to putting 
rare $B$ decays on a systematic theoretical footing,
and rightly has found its place at the
heart of many phenomenological works and fits to BSM effects. In
particular it removes most of the ambiguities of older
``$C_{7,9,10}$ + resonances'' approaches. Its primary
limitations are incalculable power corrections. There is no evidence that
these are abnormally large. Rather, the fact that they matter in the
interpretation of anomalies points to the impressive precision that
experiment has already reached (and the apparent smallness of possible
BSM effects). At the moment, there is no first-principles
method to compute power corrections. Light-cone sum rule calculations can
provide information. As they carry their own uncontrolled
systematics, it would be particularly desirable to have sum rule results
\textit{directly} for the power-suppressed terms, where possible. Combining
these with the leading-power expressions would remove most
of the systematics of either framework.
One example is the sum rule  \cite{Khodjamirian:2010vf}
for $h_+(q^2)$, which vanishes in the heavy-quark
limit. One finds an extra (ie double) power suppression
of this term at $q^2 \approx 0$, which implies an excellent
sensitivity to $C_7'$ \cite{Jager:2012uw,Jager:2014rwa}.
Data-driven approaches may be able to constrain some of the power corrections,
especially if data in very small bin-sizes becomes available for $B \to K^*
\mu^+ \mu^-$. This path has been followed to some extent in
\cite{Ciuchini:2015qxb}.

\section{Form-factors (Z.Liu \& R.Zwicky)}
\label{sec:formfactors}

Form-factors (FFs) describe the short-distance part of the transition amplitudes. 
For hadronic transitions of the type $B \to M$ at the quark level $b \to q$ 
they consist of matrix elements of the form $  \matel{M(p)}{\bar s \Gamma b}{\bar B(p_B)} $. 
For $M$ being a light meson  the FFs 
can be computed from light-cone sum rules (LCSR) and  lattice QCD at low and high momentum 
transfer $q^2 = (p_B- p)^2$ respectively. The ones most relevant to the current discussion 
of flavour anomalies are the $B \to V$ ($ V = K^*, \phi , \dots $) FFs which follow from the vector and tensor current  (cf. $O_{9,10}$ and  $O_7$ in \eqref{eq:Heff}) 
\begin{alignat}{4}
 &     \matel{K^*(p,\eta)}{\bar s \gamma^\mu (1 \mp \gamma_5) b}{\bar B(p_B)}  
 &\;=\;&   \;\;   P_1^\mu \, \V_1(q^2) &\;\pm\;& P_2^\mu \, \V_2(q^2) &\;\pm\;& P_3^\mu \,  \V_3(q^2)  \pm P_P^\mu \V_P(q^2) 
   \; ,\nonumber  \\[0.1cm]
  & \matel{K^*(p,\eta)}{\bar s iq_\nu \sigma^{\mu\nu} (1 \pm \gamma_5) b}{\bar B(p_B)} 
  &\;=\;& \;\;  P_1^\mu  T_1(q^2)   &\;\pm\;&   P_2^\mu  T_2(q^2) &\;\pm\;&  P_3^\mu  T_3(q^2) 
   \; ,
   \label{eq:ffbasis}
\end{alignat}
where we have chosen  $V = K^*$ as a matter of concreteness.
Above $P_{1,2,3,P}$ are Lorentz structures involving 
the $K^*$-meson polarisation vector $\eta$ and momenta  and the structures $\V_{1,2,3,P}$ 
are more commonly known as the $V$, $A_{2,3,0}$ FFs \cite{Straub:2015ica}. 
Below we summarise the status of these computations in LCSR \& lattice, discuss the issue of the finite width of the $K^*$-meson and the use of the equation of motion (EOM) on eliminating uncertainties. 
The FFs are fitted by a $z$-expansion, with flat priors on the coefficients, to  
 the lattice \cite{Horgan:2013hoa,Horgan:2015vla} and LCSR \cite{Straub:2015ica} data.
 The latter 
are pseudo-data generated  from the analytic LCSR computation with a Markov chain process \cite{Straub:2015ica}. 
The thereby obtained  error correlation matrix reduces the 
$10\%$-uncertainty of individual FFs   by a considerable amount,  
which is relevant for $B \to K^*\ell \ell$-decay angular distributions.

\subsection{Status of LCSR and lattice computations}
\label{sec:FFLCSR}

LCSR FFs are computed from a light-cone OPE, valid at $q^2 \leq O( m_b \Lambda) \simeq 14 \gev^2$ in an $\alpha_s$- and  twist-expansion 
resulting in convolutions of a hard kernel and light-cone distribution amplitudes (LCDA). The LCDA
 are subjected to equations of motion (EOMs) and are relatively well-known to the necessary order in the 
 conformal partial wave expansion (e.g. Gegenbauer moments).  
For  $B \to V$ ($  B_{(q,s)} \to (K^*,\phi,\rho,\omega)   $)  FFs, defined in \eqref{eq:ffbasis} the  FFs are known 
up to twist-3 at $O(\alpha_s^3)$ and twist-4 $O(\alpha_s^0)$ \cite{Ball:2004rg,Straub:2015ica}.\footnote{\label{foot:alter} Alternatively  one may use $B$-meson DA and an interpolating current for the FFs. 
E.g. \cite{Khodjamirian:2006st} for a tree-level computation with therefore slightly larger uncertainties 
with  results compatible with  \cite{Ball:2004rg,Straub:2015ica}.} State-of-the-art computations of 
$B \to K$ LCSR FFs, for which the DAs are better known because of the absence of finite width effects, can be found in \cite{Ball:2004ye,Khodjamirian:2010vf}.

Lattice QCD calculations are based on the path integral formalism in Euclidean space. 
The QCD theory is discretized on a finite space time lattice. Correlation functions, from which FFs can be extracted,
are then obtained by solving the integrals numerically
using Monte Carlo methods. Since both the noise to signal ratio of correlation functions and discretization effects increase as the momenta of hadrons increase,
lattice QCD results cover the high $q^2$ region at $\sim 15\mbox{ GeV}^2\le q^2\le q^2_{\mbox{max}}$ for $B \to V$ ($  B_{(q,s)} \to (K^*,\phi)$) FFs.
Unquenched calculations are available for 2+1-flavor  dynamical fermions~\cite{Horgan:2013hoa,Horgan:2015vla,Flynn:2016vej}
in the narrow width approximation of the vector mesons. $B \to K$ FFs can be found in~\cite{Bouchard:2013pna,Bailey:2015dka} for  2+1-flavor dynamical configurations.

\subsection{Finite width effects}

The vector meson decays via the strong force, e.g. $K^* \to K \pi$ and do therefore have a 
sizeable width and it is legitimate to ask how each formalism deals with this issue.
 
For LCSR the answer is surprisingly pragmatic in that the formalism automatically adapts to the 
experimental handling of the vector meson resonance.
In  LCSR the vector meson is described by LCDAs, which we may schematically write as
\begin{equation}
\matel{K^*(p,\eta)}{\bar s(x) \gamma_\mu  q(0)}{0} =  m_{K^*} f_{K^*} p_\mu \int_0^1 du e^{u p \cdot x} 
\phi_\parallel(u)
+ \text{higher twist} \;,
\end{equation} 
where $p = m_{K^*} \eta$, valid upon neglecting the higher twist corrections  $O(x^2,m_K^*)$, has been assumed. 
The variable $u$ is the momentum fraction of the $s$-quark in the infinite momentum frame 
and the DA $\phi_\parallel(u)$ parametrises the strength of the higher Gegenbauer moments 
(conformal spin). Since the latter contribute only about $10$-$15\%$ at the numerical level the main 
part is effectively described by the $K^*$-meson decay constant $f_{K^*}$. Hence at the pragmatic level 
the issue of finite width effects is   the same as to how well-defined the 
decay constant $f_{K^*}$ is. 

Whereas the latter can be computed from QCD sum rules or lattice modulo finite width effects it is 
advantageous to directly extract  it from experiment (e.g.  $\tau \to K^*   \nu$  and cf. 
 appendix~C in Ref.~\cite{Straub:2015ica} for a review). 
Consistency is ensured if in  both cases, leptonic 
$\tau \to (K \pi)_{l=1} \nu$ 
and hadronic decay $B \to (K \pi)_{l=1} \ell \ell$,
 the experimentalist employs 
the same fit ansatz for the resonances and  the continuous background in the p-wave ($l=1$) $K \pi$-channel. 
The transversal decay constant  $f_{K^*}^\perp$,  is  not directly accessible  in experiment. 
It is preferably taken from the ratio $f_{K^*} /f_{K^*}^\perp$  which can be computed from QCD sum rules or lattice QCD for which one would  expect finite width effects to drop out in  ratios.  
This  is a reasonable and testable hypothesis.
 In conclusion a consistent treatment of the $K^*$ in experiments  allows us to bypass a first principle definition of the $K^*$-meson 
as a pole on the second sheet and the induced error can be seen as negligible compared to the remaining uncertainty.

A fully controlled lattice QCD computation of FFs involving a vector meson needs to include scattering states. This requires much more
sophisticated and expensive calculations. In existing lattice calculations of the FFs,
the threshold effects are assumed to be small in the narrow width approximation. 
Since the $\phi$ is relatively narrow, one might expect
this approximation to be better than in the case of $K^*$. 
For $B\rightarrow D^*$ form-factors, heavy meson
chiral perturbation theory predicts 1-2\% threshold effects~\cite{Randall:1993qg,Hashimoto:2001nb} 
which is unfortunately not indicative since it is enhanced and also outside any linear regime as 
compared to the  
broad light vector meson $\Gamma_{D^*} / \Gamma_{K^*} \sim O(10^{-2})$.
An encouraging aspect is though that 
 the continuum- and chiral-extrapolated lattice FFs  obtained in~\cite{Horgan:2013hoa,Horgan:2015vla}  agree with LCSR results, upon $z$-expansion extrapolation, 
 despite the two methods having very different systematic uncertainties.
Finite width effects can only be fully assessed from a  systematic computation to which 
the Maiani-Testa no-go theorem~\cite{Maiani:1990ca} is an obstruction. 
The latter states that there is no simple relation between Euclidean ( lattice QCD) correlators
   and the transition matrix elements in Minkowski space 
   when multiple hadrons are involved in the initial or final states.
The first formalism to overcome this problem is the Lellouch-L\"uscher method~\cite{Lellouch:2000pv} which relates
matrix elements of currents in finite volume to those in the infinite volume.
Further  developments of this formalism  can be found 
in, for examples, Refs.~\cite{Hansen:2012tf,Briceno:2014uqa,Dudek:2014qha,Prelovsek:2013ela,Wilson:2014cna,Agadjanov:2016fbd}. 
Their implementations for  $B\rightarrow K\pi$-type matrix elements 
may be hoped for in the foreseeable future.

\subsection{The use of the equations of motion (projection on $B$-meson ground state)}
\label{sec:eom}

 Both in LCSR and lattice computations the $B$-meson is described by an interpolating current and it is therefore
a legitimate question  what the precision of the projection on the $B$-meson ground state is.\footnote{In  both cases this is in practice achieved by an exponential suppression of the higher states. 
If one was able to compute the correlation function exactly in Minkowski space then one could use the LSZ-formalism. Hence the method used can be seen as approximate methods to the latter.} 
Below we argue that the EOMs  improve the situation.  

The vector and tensor operators entering \eqref{eq:ffbasis} are related by EOMs
\begin{alignat}{4}
\label{eq:eom}
& \matel{K^*}{ i \partial^\nu (\bar s i \sigma_{\mu \nu}  b) }{B}   &\;=\;&
   - (m_s \pm m_b)   \matel{K^*}{ \bar s \gamma_\mu  b }{B} &\;+\;& i \partial_\mu 
    \matel{K^*}{ (\bar s  b) }{B} \;,
   &\;-\;& 2 \matel{K^*}{  \bar s i \!\stackrel{\leftarrow}{D}_{\mu}  b}{B} \nonumber \\[0.1cm]
   & \quad \sim T_1(q^2)  &\;=\;& \quad \sim \V_1(q^2) &\;+\;&  \quad \sim 0 &\;+\;& \quad \sim  {\cal D}_1(q^2)
   \end{alignat}
  where the derivative  term defines a new FF ${\cal D}_1(q^2)$ in analogy to $T_{1}(q^2)$.  
  The EOM \eqref{eq:eom} are exact and have to be obeyed  
   and therefore can be used as a non-trivial check  for any computation. 
  This has been done in \cite{Straub:2015ica} at the tree-level up to twist-4 and 
  for the $Z$-factors describing the renormalisation of the composite operators entering   the EOM \eqref{eq:eom}.
  Furthermore 
  at the level of the effective Hamiltonian this term is redundant, since the EOMs have  been used in reducing the basis of operators. 
  Eq.\eqref{eq:eom} therefore defines a relation between 3 FFs of which one is redundant which does
  not appear helpful at first. The use comes from the hierarchy ${\cal D}_1 \ll T_1,\V_1$ which therefore 
  constrains the vector in terms of the tensor FF \cite{Hambrock:2013zya,Straub:2015ica}.  
 At the level of the actual computation this allows us to control the correlation of the continuum threshold parameters which are a major source of uncertainty for the individual FFs. The argument is that 
 if those parameters were to differ by a sizeable amount for the vector and tensor form factors then 
 the exactness of \eqref{eq:eom} would impose a huge shift for the corresponding parameters of the
 derivative FF corresponding to an absurd violation of semi-global quark hadron duality.  
 Since ${\cal D}_1$ seems to be a FF with normal convergence in $\alpha_s$ and the twist-expansion this possibility seems absurd and therefore supports the validity of the argument. 
It should be mentioned that the crucial hierarchy ${\cal D}_1 \ll T_1,\V_1$ can be traced back 
  to the large energy limit \cite{Charles:1998dr}.
For more details and a plot illustrating the validity of this argument over the $q^2$-range we refer 
the reader to references \cite{Hambrock:2013zya,Straub:2015ica} and Fig.1 in \cite{Straub:2015ica}.

 On the lattice the projection on the ground state would be perfect if one could go to infinite Euclidean 
time and if there was no noise. In practice simulations are done at a finite $t$-interval 
with some noise and this sets some limitations on the projection. On the positive side these aspects are not the main sources of uncertainties in current LQCD calculations (smearing and other methods are being used to improve the projection)
and
are improvable with more computer power. 
It is conceivable  that the EOMs can be used for the lattice in correlating the projection 
of FFs entering the same EOM. 

\section{The relevance of charm contributions (R.Zwicky)}
\label{sec:charm}

\subsection{The LCSR framework}

The long distance contributions can be computed with LCSR as an alternative to the 
heavy quark framework discussed in section \ref{sec:heavy}.  
The  basic long distance topologies
are  well-known and include the chromomagnetic operator $O_8$ \eqref{eq:Heff}, the weak 
annihilation (WA), quark-spectator loop scattering (QQLS) 
and the charm contributions. The latter three  originate 
from the four quark operators $O_{1 ..6}$ in \eqref{eq:Heff}. The charm contributions 
are presumably the most relevant  since they are not CKM suppressed 
with sizeable Wilson coefficients and are the   
subject of a longer discussion in  Sec.~\ref{sec:charm2}. 
For $B \to K^{(*)} \ell \ell$ and other final state hadrons the 
$O_8$ and  WA contribution have been computed in \cite{Dimou:2012un}  and 
\cite{Lyon:2013gba} respectively. Both  come with strong phases and the first 
one is found to be small whereas the second one is rather sizeable albeit CKM suppressed. 
The quark spectator scattering diagrams have been computed in a hyprid approach of
heavy quark expansion with infrared divergences removed by an LCSR computation \cite{Lyon:2013gba}.

Whereas we agree that the heavy quark expansion provides a systematic framework 
in the $1/m_b$-limit for regions outside the resonance regions it has to be seen that 
in practice  
these conditions are not always met for $b  \to s \ell \ell$ observables.
Since LCSR are not dependent on the $1/m_b$-limit 
they do provide an alternative to estimate these effects. 
In Sec.~\ref{sec:charm2} we  argue
that the $1/m_b$ vertex corrections 
might well be sizeable, contributing to potential tension in $b \to s \ell \ell$-observables.  
Resonant effects can be systematically parameterised and fitted for but their inclusion into 
predictions requires further thought and work when a sizeable continuum contributions is present.

\subsection{The charm contribution}
\label{sec:charm2}

\begin{figure}[h]
\centering
\includegraphics[width=9cm,clip]{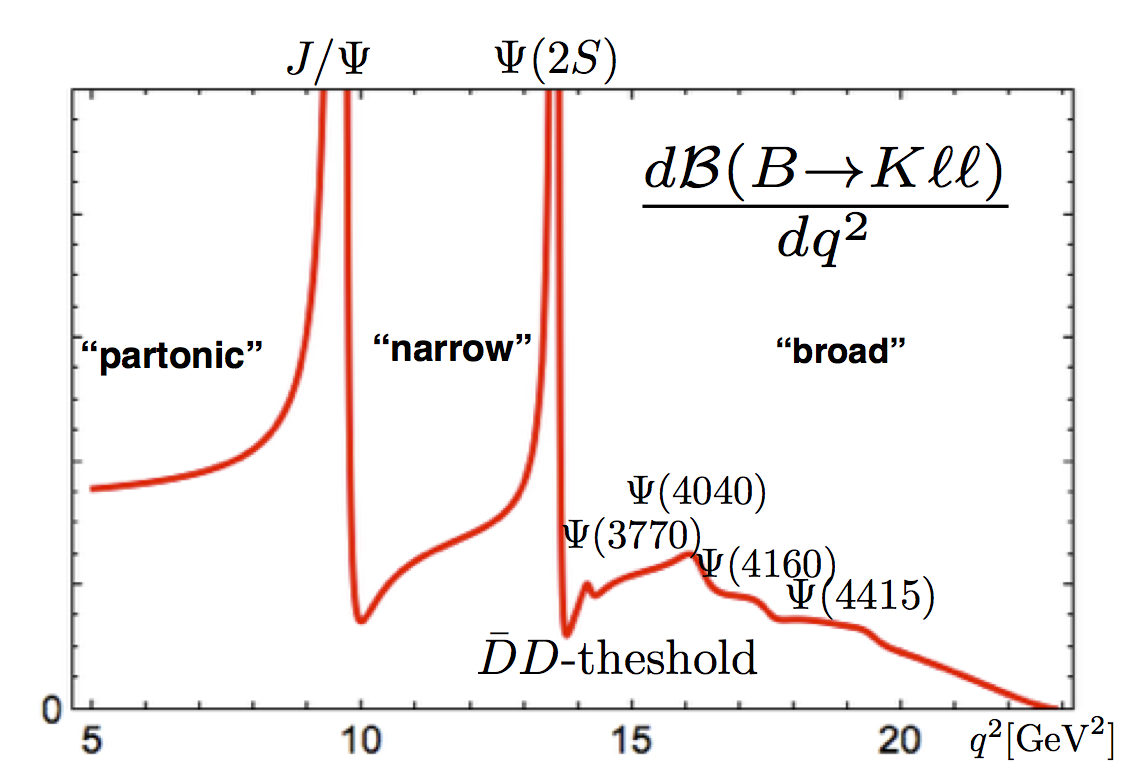}
\caption{\small Illustration of the spectrum in $q^2$, the lepton pair invariant momentum, for 
the $B \to K \ell \ell$ branching fraction. The three regions referred to in the text are the 
``below resonance region" (partonic, low-$q^2$), ``narrow resonance region" and the ``broad resonance region" (high-$q^2$).}
\label{fig-c}       
\end{figure}

In FCNC processes of the type  $B  \to K^{(*)} \ell \ell$, the subprocess 
$B \to  K^{(*)}(\bar c c \to \gamma^* \to \ell \ell) $ is numerically 
relevant since it proceeds at tree-level.  Hence the Wilson coefficients 
of operators of the type  $O^{(c)}_{1,2}$ \eqref{eq:Heff} in the effective Hamiltonian are sizeable.
The other relevant aspect is that the four momentum invariant, $q^2 \in [4 m_\ell^2,  (m_B - m_{K^{(*)}})^2]$, of the lepton-pair 
takes on values in the region of charmonium resonances. Hence the process is sensitive to
resonances, with photon quantum numbers $J^{PC} = 1^{--}$, and one can therefore not  
entirely  rely on a  partonic picture. It is customary to divide the $q^2$-spectrum  into three regions (cf.Fig.~\ref{fig-c}).  A region sufficiently well-below the first charmonium resonance at
 $q^2 = m_{\jpsi}^2 \simeq 9.6 \gev^2$, the region of the two narrow charmonium resonances $\jpsi$ and 
 $\psi(2S)$ and the region of broad charmonium states above the  $\bar DD$-threshold 
 $q^2 \simeq 14 \gev^2$. 
The crucial question is  when and how partonic methods are applicable in the hadronic charmonium region. We discuss them below in the order of tractability in the partonic picture. 

In the ``below-charmonium region"   
 partonic methods (i.e. $\al_s$-expansion ) are expected to be applicable. Since this is the region 
where the particles are fast in the $B$-rest frame, the physics  can be described within a light-cone formalism. 
The order $O(\al_s^0)$ contribution is equivalent to naive factorisation (NF) which means 
that the amplitude factorises as follows
\begin{equation}
\label{eq:naiveFAC}
{\cal A}[B \to K^{(*)}) \ell \ell]\Big|^{\text{NF}}_{O^{(c)}_{1,2}}
 \sim h(q^2) F^{B \to K^{(*)}} (q^2) \;,
\end{equation}
where $F^{B \to K^{(*)}}(q^2)$ stands 
for the relevant  form factor (FF)  combination and $h(q^2)$ is the vacuum polarisation due to the charm-part 
of the electromagnetic current.  
  At this formal level the light-cone aspect is only present in 
that the preferred methodology  for evaluating the FFs are light-cone methods (cf. Sec.~\ref{sec:FFLCSR}). 
 Corrections of order $O(\al_s)$ are expected to be sizeable since the leading order
contribution is of the colour-suppressed Wilson coefficient combination. 
Simple factorisation formulae only hold in the $1/m_b$-limit which include hard spectator and 
vertex corrections \cite{Bosch:2001gv,Beneke:2001at} with the latter borrowed from inclusive computations \cite{Asatrian:2001de}.  The $O(\al_s/m_b)$ corrections  can be done within an extended 
LCSR approach with either a  $B$- or $K^{(*)}$-meson LCDA. Only partial results exist in that 
the vertex and hard spectator corrections have not been evaluated in either of these methods.
The third contribution, emission of the gluon into the LCDA, have been done for $K^{(*)}$- and $B$-meson  \cite{Ball:2006eu,Muheim:2008vu} (computations 
for $q^2=0$ only) and \cite{Khodjamirian:2010vf} respectively.
The reason that the vertex and hard spectator corrections  
have not been performed in either  approach is that they are technically demanding. 
If one insists on verifying the dispersion relation involved this demands 
an evaluation of a two-loop graph with five scales and an integration over
one LCDA-parameter. 
The size of these contributions is therefore unknown beyond the heavy quark limit
and might well be sizeable enough to explain the current picture of deviations even 
within a partonic approach. 
A completion of this program is therefore desirable.

The broad charmonium region (see Fig.~\ref{fig-c} on the right) is characterised by a considerable 
interference of the short-distance and charmonium long-distance part.  
The charmonium resonances $\psi(3770)$, $\psi(4040)$, $\psi(4160)$ and $\psi(4415)$ are broad 
because of they can decay via the strong force into  $\bar DD$-states. 
The resonances are necessarily accompanied by 
a continuum of $\bar DD$-states as is the case in $e^+ e^- \to hadrons$.  
\cite{Lyon:2014hpa}.  A sketch of a realistic parametrisation is given by\footnote{More realistic 
parameterisation 
differ in two aspects. Firstly, they only parameterise 
the discontinuity and the remaining 
part is obtained by a dispersion relation e.g.  $e^+ e^- \to hadrons$  \cite{Ablikim:2007gd} and 
$B \to K \ell \ell$  \cite{Lyon:2014hpa}. Second they include energy dependent width effects and interferences of the overlapping broad resonances. This has been successfully done in $e^+ e^- \to hadrons$  \cite{Ablikim:2007gd} and 
adapted to $B \to K \mu \mu$ in \cite{Lyon:2014hpa}. The inclusion of interference effects reduces 
the $\chi^2/\text{d.o.f.}$ from $\simeq1.4$ to $\simeq 1$. Crucially the degree of model dependence is 
justified by the goodness of the fit.}
\begin{equation}
\label{eq:charm}
{\cal A}[B \to K^{(*)}) \ell \ell]\Big|_{O^{(c)}_{1,2}}   \simeq 
\sum_{\psi = \jpsi, \psi(2S),\psi(3770),.. }   \frac{r_{\psi}}{q^2 - m_{\psi}^2 + i m_\psi \Gamma_\psi}  
 +  f_{\bar cc}(q^2)  \;,
\end{equation}
where the residues $r_\psi$ and the non-resonant $\bar cc$-continuum function $f_{\bar cc}$
are the unknowns which usually have to be determined   experimentally.
For $e^+ e^- \to hadrons$, which correspond effectively to  NF \eqref{eq:naiveFAC}, 
$r_\psi^{\text{NF}} \sim   \Gamma(\psi \to \ell \ell)F^{B \to K^{(*)}} (m_\psi^2) > 0$ 
and   $\text{Im}[f^{\text{NF}}_{\bar cc}(q^2)  ] > 0$  
for which successful parametrisations are easily found and the real part follows from  
a dispersion relation \cite{Ablikim:2007gd,Lyon:2014hpa}. 
Going beyond factorisation involves determining the factors $\eta_\psi$ and the continuum function $ f_{\bar cc}(q^2)$
\begin{equation}
\eta_\psi \equiv   |\eta_\psi| e^{i \de_{\psi}} =   \frac{r_\psi}{r_\psi^{\text{NF}}}    \;, \qquad f_{\bar cc}(q^2)\;.
\end{equation}
The strong-phase  $ \de_{\psi} $ is defined 
relative to the  FF contribution. 
The determination of $ f_{\bar cc}(q^2)$ is a difficult task since the $q^2$-dependence is 
 similar to the one of the short distance contributions. On the other hand the shape of the resonances is 
 very distinct (e.g Fig.~\ref{fig-c}) and can  therefore be fitted unambiguously. 
 This has been done in  \cite{Lyon:2014hpa}
using the LHCb-data \cite{LHCb-PAPER-2013-039}. A surprisingly good fit is obtained by 
$\eta_{\tilde{\psi}} \simeq -2.55$ and $\de_{\tilde{\psi}} \simeq 0$
with 
$\tilde{\psi} \in \{\psi(3770),\psi(4040),\psi(4160),\psi(4415) \}$.\footnote{Cf. Ref.~\cite{Lyon:2014hpa} 
for more realistic fits, with variable individual complex residues, with slightly improved $\chi^2/\text{dof}$.
These findings have recently been confirmed 
by the LHCb collaboration \cite{Aaij:2016cbx}.}
 One concludes that NF is badly broken and that the charm 
might impact on the low $q^2$-spectrum \cite{Lyon:2014hpa}. 
For example, even in the narrow width 
approximation the non-local part of the resonances \eqref{eq:charm} only decays 
as $1/q^2$ away from its centre. 

These results bring the focus to the ``narrow charmonium region".
Whereas the absolute values 
$|\eta_{\jpsi}| \simeq 1.4$ and $|\eta_{\psi(2S)} | \simeq 1.8$ are known from the decay rates 
$\Gamma(B \to K^{(*)} \psi)$ the phases $\de_{\jpsi,\psi(2S)}$ are unknown. 
The need to extract these from experiment has been emphasised and suggested in  
\cite{Lyon:2014hpa} and recently been performed by LHCb collaboration \cite{Aaij:2016cbx}. 
Unfortunately, so far,  the solutions show a four-fold ambiguity 
$(\de_{\jpsi} | \de_{\psi(2S)} ) \simeq  ( 0, \pi | 0,\pi   )  $. Whereas 
this is an important result  one would 
hope that this ambiguity can be resolved with more data in the future.

After this excursion let us discuss the practicalities for the low-$q^2$ (``below resonance region")
and the high-$q^2$ (``broad resonance region") regions where phenomenologists and experimentalists 
compare predictions to measurements in the hope of seeing physics beyond the Standard Model. 
It would be desirable to obtain a coherent picture of the partonic and hadronic descriptions in both 
of these regions in order to validate the approaches.  Both cases need more work.
At low-$q^2$, as  discussed above, the partonic contributions are not very complete. 
For example, the potentially sizeable vertex corrections of the charm loop are unknown beyond the $1/m_b$-limit.
There is no universal or well-understood pattern for estimating the size of the $1/m_b$-corrections. 
For example the FF  $1/m_b$-corrections are around $10\%$, for the $B$-meson decay constant $f_B$ they are $30\%$ 
 whereas for the 
$O_8$-matrix elements (chromomagnetic operator) they are $50\%$ and more  \cite{Dimou:2012un}. 
As for the hadronic data, further work is needed in order to determine  the strong phases 
$\de_{\jpsi,\psi(2S)}$ as well as the continuum function $f_{\bar cc}(q^2)$. 
Before moving on it should be mentioned that the hadronic fits should and will be extended 
from $B \to K \ell \ell$ to $B \to K^* \ell \ell$ by the LHCb-collaboration.
In the high-$q^2$ region a partonic picture has been advocated, 
known as the high-$q^2$ OPE, where one resorts 
to an expansion in $1/m_b$ and $1/\sqrt{q^2}$ supplemented by charm-contributions 
\cite{Grinstein:2004vb,Beylich:2011aq}.  The initial idea was   
to include charm contributions in an $\al_s$-expansion,  relying on cancellations when 
averaged over large enough bins.\footnote{
The gateway to quark-hadron duality are dispersion relations valid at the level of amplitudes.
Averaging at the decay rate level is only valid if the rate can be written as an amplitude which
is the case for inclusive modes such as $e^+ e^- \to hadrons$. 
Averaging over the entire $B \to K^{(*)} \ell \ell$-rate, which includes the narrow resonances, 
fails by several orders of magnitudes as discussed and clarified in  \cite{Beneke:2009az}.}
As an estimate of these type of  quark-hadron duality violations for the broad resonances, 
naive factorisation (i.e. $e^+ e^- \to hadrons$)  was taken as  guidance \cite{Beylich:2011aq}, 
which suggests an effect of the order of  $2\%$. 
One is then faced with the a posteriori fact that the actual data 
\cite{LHCb-PAPER-2013-039} show effects in the region of $10\%$ 
\cite{Lyon:2014hpa}.\footnote{For angular observable effects might look though 
more favourable in two aspects partly at the cost of sensitivity to new physics.
Firstly, in the limit of no right-handed currents and 
constant $\eta_{\tilde{\psi}}$ resonance effects drop out in a few angular observables 
\cite{Hambrock:2013zya}. Second at the kinematic endpoint $q^2 = (m_B - m_{K^{(*)}})^2$ 
the observables approach exact values based on Lorentz-covariance only (i.e. valid in any model and approximation which respects Lorentz invariance)  and show a certain 
degree of universality  near the endpoint region \cite{Hiller:2013cza}.}
Hence supplementing the high-$q^2$ OPE with  a hadronic representation 
\eqref{eq:charm} seems attractive.  
The bottleneck is though the determination of the  the continuum function $ f_{\bar cc}(q^2)$ \eqref{eq:charm}.
A promising direction could be the experimental investigation of the $B \to \bar DD K^{(*)}$ modes.

\subsubsection{Discussion and summary}

Clarifying the role of the charm will remain an outstanding task before 
 angular anomalies of the $B\to K^* \ell \ell$-type and branching fraction deviations, 
 not related to violation of lepton flavour universality (LFU), can be considered
 to be physics beyond the 
 Standard Model. 
 Progress can be made by computing 
 charm contributions consistently in one approach, complemented by  
 more refined experimental data allowing us to extract the relevant information. 
 Hopefully this will  pursued by experimentalists and theorists. 
 
 Finally,  a few brief comments on LFU-violation,  the possibility of 
 new physics in charm and right-handed currents. 
 Clarifying the role of LFU-violating observables, e.g. $R_K$ \eqref{eq:RK}, is of importance for global fits. 
 QED corrections, which can give rise to $O(\al_{\text{QED}} \ln^2(m_\mu/m_e))$-deviations  from
 $R_K=1$,   can be diagnosed by higher moments \cite{Gratrex:2015hna}.
 The dominant real radiation effect  has been estimated  in \cite{Bordone:2016gaq} 
 and crucially the effect deviates by  $O(1\%)$  
 as compared to the PHOTOS-implementation~\cite{Golonka:2005pn} used by the LHCb-analysis. 
 Migration of narrow charmonium events into lower $q^2$-region 
 by misidentification of hadronic final states 
 \cite{Gratrex:2015hna} has been  checked by the LHCb-collaboration.  
 The fact that charm contributions can mimic shifts in $C_9$, or any other amplitude with photon quantum numbers, has been emphasised  in \cite{Lyon:2014hpa} 
 motivated by  the results on the broad   charmonium resonances in $B \to K \ell \ell$.  In this work 
 it was suggested to diagnose this effect with $q^2$-dependent fits in various $b \to s \ell \ell$-channels,  
 and the question as to whether 
 the anomalies  could  partially be due to new physics in charm (e.g. $\bar b  c \bar c s$-operators) was raised. 
 The former idea was applied to $B \to K^* \ell \ell$ at low $q^2$ 
 for the first time in ~\cite{Altmannshofer:2015sma},  
and the possibility of charm in new physics was recently investigated systematically including RG-evolution
and constraints (e.g. $B_s$-mixing) in  \cite{Jager:2017gal}. 
A closely related area  where charm contributions are of importance is  the search for right-handed currents. 
The V-A structure of the weak interactions and the small ratio $m_s/m_b$ suppress amplitudes 
with right-handed quantum numbers that can, for instance, be measured in time-dependent CP-asymmetries \cite{Atwood:1997zr}.
Charm contributions, of higher-twist, are a non-perturbative background to these measurements, 
whose understanding is important. 
LHCb's first time measurement of the $B _s \to \phi \gamma$ time-dependent CP-asymmetry 
${\cal A}_\Delta \simeq -0.98 (50)(20)$ \cite{Aaij:2016ofv} comes with a large uncertainty but also with 
a large deviation from the SM prediction  ${\cal A}_\Delta \simeq 0.047 (28)$ \cite{Muheim:2008vu},
which allows for speculations in various directions.

\section{Global fits (L. Hofer)}
\label{sec:global}

As reported in Sec.~\ref{sec:experiment}, experimental results on $B\to K^*\mu^+\mu^-$, $B_s\to\phi\mu^+\mu^-$ and
$R_K=Br(B\to K\mu^+\mu^-)/Br(B\to Ke^+e^-)$ show deviations from the SM at the $2-3\sigma$ level.
Although none of these tensions is yet significant on its own, the situation is quite intriguing as
the affected decays are all mediated by the same quark-level transition $b\to s\ell^+\ell^-$ 
and thus probe the same high-scale physics. A correlated analysis of these channels can shed light 
on the question whether a universal new-physics contribution to $b\to s\ell^+\ell^-$
can simultaneously alleviate the various tensions and lead to a significantly improved global 
description of the data.

At the energy scale of the $B_{(s)}$ decays, any potential high-scale new physics 
mediating $b\to s\ell^+\ell^-$ transitions can be encoded
into the effective couplings $C_{7,9,10}^{(\prime)}$ multiplying the operators
\begin{eqnarray}
  &&\mathcal{O}_9^{(\prime)}=\frac{\alpha}{4\pi}[\bar{s}\gamma^\mu P_{L(R)}b]
   [\bar{\mu}\gamma_\mu\mu],\hspace{1cm}
  \mathcal{O}_{10}^{(\prime)}=\frac{\alpha}{4\pi}[\bar{s}\gamma^\mu P_{L(R)}b]
   [\bar{\mu}\gamma_\mu\gamma_5\mu],\nonumber\\
  &&\mathcal{O}_7^{(\prime)}=\frac{\alpha}{4\pi}m_b[\bar{s}\sigma_{\mu\nu}P_{R(L)}b]F^{\mu\nu},
\end{eqnarray}
where $P_{L,R}=(1 \mp \gamma_5)/2$ and $m_b$ denotes the $b$ quark mass.
Whereas the above-mentioned semi-leptonic decays are sensitive to the full set $C_{7,9,10}^{(\prime)}$ of
effective couplings, the decay $B_s\to\ell^+\ell^-$ only probes $C_{10}^{(\prime)}$ and 
$B\to X_s\gamma$, $B\to K^*\gamma$ set constraints on  $C_{7}^{(\prime)}$.
Note that additional scalar or pseudoscalar couplings $C_{S,S',P,P'}$ cannot address the tensions in 
exclusive semi-leptonic $B$ decays since their contributions are suppressed by small lepton masses.
  
\begin{table}
\scriptsize
\centering
\begin{minipage}{0.45\linewidth}
\begin{tabular}{@{}crccc@{}}
\toprule[1.6pt] 
Coefficient & Best fit & 1$\sigma$ & Pull$_{\rm SM}$ \\ 
 \midrule 
 $C_7^{\rm NP}$ & $ -0.02 $ & $ [-0.04,-0.00] $ &  1.2 \\[1mm] 
 \boldmath$C_9^{\rm NP}$ & \boldmath$ -1.09 $ & $ [-1.29,-0.87] $ &  \bf 4.5 \\[1mm] 
 $C_{10}^{\rm NP}$ & $ 0.56 $ & $ [0.32,0.81] $ &  2.5 \\[1mm] 
 $C_{7}^{\prime\rm NP}$ & $ 0.02 $ & $ [-0.01,0.04] $ &  0.6 \\[1mm] 
 $C_{9}^{\prime\rm NP}$ & $ 0.46 $ & $ [0.18,0.74] $ &  1.7 \\[1mm] 
 $C_{10}^{\prime\rm NP}$ & $ -0.25 $ & $ [-0.44,-0.06] $ &  1.3 \\[1mm] 
 $C_9^{\rm NP}=C_{10}^{\rm NP}$ & $ -0.22 $ & $ [-0.40,-0.02] $ &  1.1 \\[1mm] 
 \boldmath$C_9^{\rm NP}=-C_{10}^{\rm NP}$ & \boldmath$ -0.68 $ & $ [-0.85,-0.50] $ &  \bf 4.2 \\[1mm] 
 \boldmath$C_9^{\rm NP}=-C_{9}^{\prime\rm NP}$ & \boldmath$ -1.06 $ & $ [-1.25,-0.85] $ &  \bf 4.8 \\[1mm] 
\bottomrule[1.6pt] 
\end{tabular}
\end{minipage}
\begin{minipage}{0.5\linewidth}
\begin{tabular}{@{}lccc@{}}
\toprule[1.6pt] 
\hspace{1cm}Fit & $C_{9\ \rm Best fit}^{\rm NP}$ & 1$\sigma$ & Pull$_{\rm SM}$ \\ 
 \midrule 
 All $b\to s\mu\mu$ & $ -1.09 $ & $ [-1.29,-0.87] $ &  4.5  \\[1mm]  
 $b\to s\mu\mu$ without $q^2\in[6,8]$ & $ -0.99 $ & $ [-1.23,-0.75] $ &  3.8 \\[1mm]
 $b\to s\mu\mu$ large recoil & $ -1.30 $ & $ [-1.57,-1.02] $ &  4.0 \\[1mm] 
 $b\to s\mu\mu$ low recoil & $ -0.93 $ & $ [-1.23,-0.61] $ &  2.8 \\[1mm]  
 Only $B\to K\mu\mu$ & $ -0.85 $ & $ [-1.67,-0.20] $ &  1.4 \\[1mm] 
 Only $B\to K^*\mu\mu$ & $ -1.05 $ & $ [-1.27,-0.80] $ &  3.7 \\[1mm] 
 Only $B_s\to \phi\mu\mu$ & $ -1.98 $ & $ [-2.84,-1.29] $ &  3.5 \\[1mm]
\bottomrule[1.6pt] 
\end{tabular}
\end{minipage}
\normalsize
\caption{Left: best-fit point, 1$\sigma$ region and SM-pull for 1-parameter fits allowing new physics only in one 
of the couplings $C_{7,9,10}^{(\prime)}$. Right: fits of a new-physics contribution to the effective coupling $C_9$ 
using different subsets of the experimental data as input. Results are taken from Ref.~\cite{Descotes-Genon:2015uva}\label{tab:fitres}}
\end{table}

Various groups have performed fits of the couplings $C_{7,9,10}^{(\prime)}$ to the 
data~\cite{Descotes-Genon:2013wba,Altmannshofer:2013foa,Altmannshofer:2014rta,Descotes-Genon:2015uva,Hurth:2016fbr}.
The obtained results are in mutual agreement with each other and confirm the observation, 
pointed out for the first time in Ref.~\cite{Descotes-Genon:2013wba} on the basis of the 2013 data,
that a large negative new-physics contribution $C_9$ yields a fairly good description of the data. This is illustrated in 
Tab.~\ref{tab:fitres}, where selected results from Ref.~\cite{Descotes-Genon:2015uva} for one-parameter fits of the couplings
$C_{7,9,10}^{(\prime)}$ are displayed. Apart from the best-fit point together with the $1\sigma$ region, the tables feature the 
SM-pull of the respective new-physics scenarios. This number quantifies by how many sigmas the best fit point is preferred
over the SM point $\{C_{i}^{\rm NP}\}=0$ and thus measures the capacity of the respective scenario to
accommodate the data. The table on the left demonstrates that a large negative new-physics contribution $C_9$
is indeed mandatory to significantly improve the quality of the fit compared to the SM. It is particularly encouraging 
that the individual channels tend to prefer similar values for $C_9$, as shown in the table on the right.

The results of the fit are quite robust with respect to the hadronic input and the employed methodology. This can be seen
from the good agreement between the results of the analyses AS~\cite{Altmannshofer:2014rta} and DHMV~\cite{Descotes-Genon:2015uva}
which use approaches that are complementary in many respects:
\begin{itemize}
  \item AS choose the observables $S_i$ as input for the fit to the angular distributions of $B\to K^*\ell^+\ell^-$ and $B_s\to\phi\ell^+\ell^-$
  and restrict their fits in the region of large $K^*$-recoil to squared invariant dilepton masses $q^2\in[0,6]$\,GeV$^2$.
  DHMV, on the other hand, choose the observables $P_i^{(\prime)}$, which feature a reduced sensitivity to the non-perturbative form factors,
  and include all bins up to $q^2=8$\,GeV$^2$.
  \item AS use LCSR form factors from Ref.~\cite{Straub:2015ica}, while DHMV mainly resort to the LCSR form factors from Ref.~\cite{Khodjamirian:2010vf}.
  \item In the analysis of AS, correlations among the form factors are implemented on the basis of the LCSR calculation~\cite{Straub:2015ica},
  whereas in the analysis of DHMV they are assessed from large-recoil symmetries supplemented by a sophisticated estimate of 
  symmetry-breaking corrections of order $\mathcal{O}(\Lambda/m_b)$. The pros and cons of these two methods complement each other: The first approach 
  provides a more complete access of correlations at the price of a dependence on and the limitation to one particular LCSR calculation~\cite{Straub:2015ica}
  and its intrinsic model-assumptions. The second approach determines the correlations in a model-independent way from first principles but needs to 
  rely on an estimate of subleading non-perturbative $\Lambda/m_b$ corrections. 
\end{itemize}
\begin{figure}
\begin{center}
\begin{minipage}{0.35\linewidth}
 \includegraphics[width=0.95\linewidth]{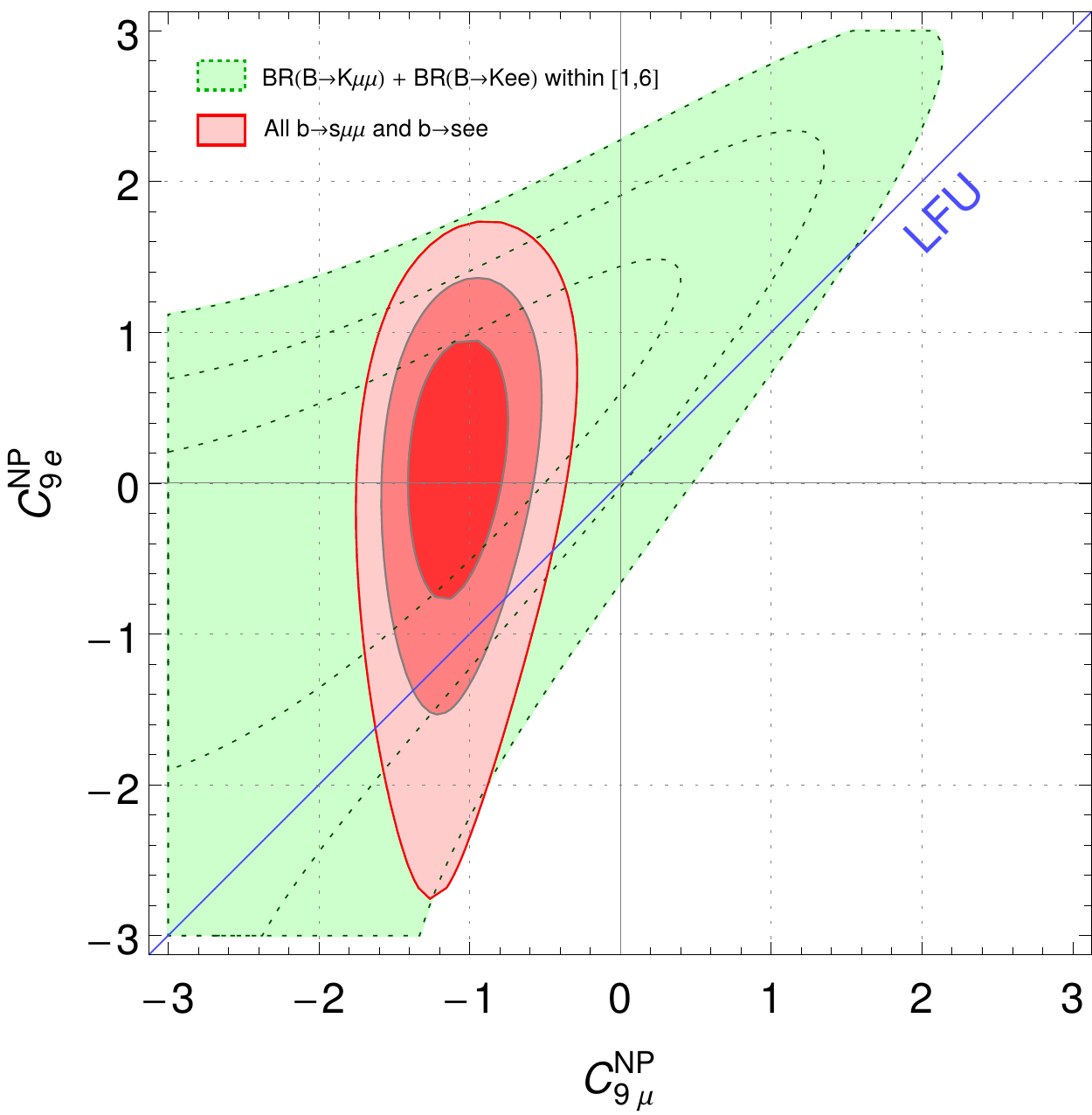}
\end{minipage}
\begin{minipage}{0.55\linewidth}
  \includegraphics[width=1.1\linewidth]{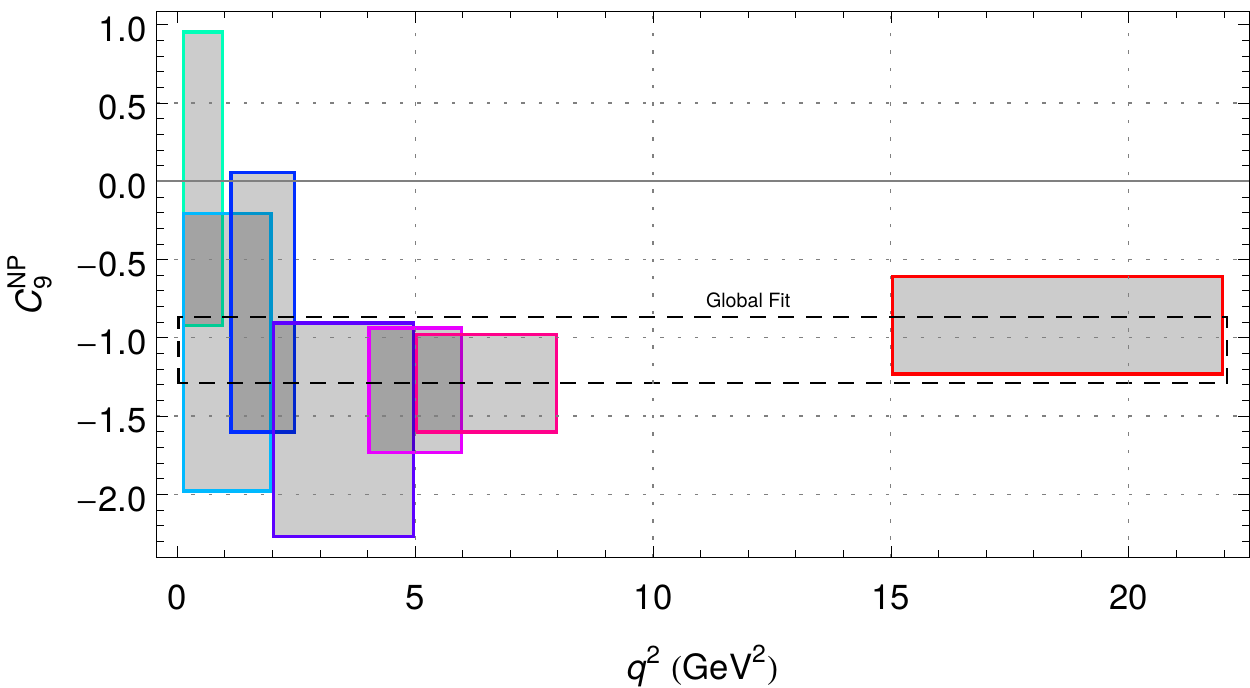}
\end{minipage}\hspace{0.08\linewidth} 
\end{center}
\caption{Left: Fit allowing for LFUV by means of independent coefficients $C_{9\;\mu}^{\rm NP}$ and $C_{9\;e}^{\rm NP}$.
Right: Bin-by-bin fit of the one-parameter scenario with a single coefficient $C_9^{\rm NP}$. Reproduced from Ref.~\cite{Descotes-Genon:2015uva}.
\label{fig:BinFit}}
\end{figure}

The measurement of $R_K\neq 1$ hints at a possible violation of lepton-flavour universality (LFUV) and suggests a
situation where the muon- and the electron-components of the operators
$C_{9,10}^{(\prime)}$ receive independent new-physics contributions $C_{i\;\mu}^{\rm NP}$ and
$C_{i\;e}^{\rm NP}$, respectively. In Fig.~\ref{fig:BinFit} on the left we display the result for the two-parameter fit to the coefficients
$C_{9\;\mu}^{\rm NP}$ and $C_{9\;e}^{\rm NP}$. The plot is taken from Ref.~\cite{Descotes-Genon:2015uva}, similar results are obtained in 
Refs.~\cite{Altmannshofer:2014rta,Hurth:2016fbr}. The fit prefers an electron-phobic scenario with new physics coupling to $\mu^+\mu^-$ but not to $e^+e^-$. Under this hypothesis,
that should be tested by measuring $R_{K^*,\phi}$ as well as lepton-flavour sensitive angular 
observables~\cite{Capdevila:2016ivx}~\footnote{Very recently, Belle has presented a separate measurement~\cite{Wehle:2016yoi}
of $P_5^\prime$ in the muon and electron channels. While the muon channel exhibits a 2.6$\sigma$ deviation with respect to the SM
prediction in good agreement with the LHCb measurement, the electron channel agrees with the SM expectation at 1.3$\sigma$.},
the SM-pull increases by $\sim 0.5\sigma$ compared to the value in Tab.~\ref{tab:fitres} for the lepton-flavour universal scenario, except for the scenario 
with $C_9^{\rm NP}=-C_{9}^{\prime\rm NP}$ where the value remains unchanged due to the absence of any contribution to $R_K$.

The fact that it is primarily the variation of the coefficient $C_9$ which is responsible for solving the anomalies unfortunately spoils 
an unambiguous interpretation of the fit results in terms of new physics.
The reason is that precisely this effective coupling can be mimicked by non-perturbative charm loops, as discussed in Sec.~\ref{sec:charm}.
However, whereas these non-local effects are expected to introduce a non-trivial dependence on the squared invariant mass $q^2$ of the lepton pair, 
a high-scale new-physics solution would necessarily generate a $q^2$ independent $C_9^{\rm NP}$.
A promising strategy to resolve the nature of a potential non-standard contribution to the effective coupling $C_9$ thus consists in the
investigation of its $q^2$ dependence. To this end, two different methods have been pursued so far: 
In Ref.~\cite{Altmannshofer:2015sma}, the authors performed a bin-by-bin fit of $C_9$ to check whether results 
in different bins were consistent with each other under the hypothesis of a $q^2$-independent $C_9$.
Their conclusion, which was later confirmed also in Ref.~\cite{Descotes-Genon:2015uva} with the plot shown in Fig.~\ref{fig:BinFit} on the right,
was that there is no indication for a $q^2$-dependence, though the situation is not conclusive due to the large uncertainties in the single bins.

An alternative strategy to address this question has been followed recently in Refs.~\cite{Ciuchini:2015qxb,Capdevila:2017ert} where a direct fit of the $q^2$-dependent 
charm contribution $C_9^{c\bar{c}\;i}(q^2)$ to the data on $B\to K^*\mu^+\mu^-$ (at low $q^2$) has been performed under the hypothesis of the absence of new physics.
The results are in agreement with the findings from Fig.~\ref{fig:BinFit}: in Ref.~\cite{Capdevila:2017ert} it was shown that the inclusion of 
additional terms parametrising a non-trivial $q^2$-dependence does not improve the quality of the fit. On the other hand, current precision of the 
experimental data does not allow to exclude non-zero values for these terms.
 
In certain scenarios, a $q^2$-dependent contribution to $C_9$ can also have its origin at high energy scales: new physics mediating 
$b\to sc\bar{c}$ transitions would induce a $q^2$-dependent contribution to $b\to s\ell^+\ell^-$ at the one-loop level in the effective theory. 
This possibility was proposed for the first time in Ref.~\cite{Lyon:2014hpa}, while a phenomenological analysis taking into account constraints
from $B_s-\bar{B}_s$ mixing and $B\to X_s\gamma$ was performed recently in Ref.~\cite{Jager:2017gal}. Note that a charm-loop 
contribution to $b\to s\ell^+\ell^-$, whether from high-scale new physics or from low-energy QCD dynamics, always conserves lepton flavour 
and thus could not account for deviations in $R_K$ or other LFUV observables.

\section{BSM interpretation (L. Hofer)}
\label{sec:bsm}

As we have seen in the previous section, the observed anomalies in $b\to s\ell^+\ell^-$ decays show a coherent picture
and allow for a solution at the level of the effective Hamiltonian by NP contributions to the operators $\mathcal{O}^{(\prime)}_{9,10}$. 
At tree level, contributions to these operators can be mediated by exchange of a heavy neutral 
vector-boson $Z^\prime$ 
(e.g.~\cite{Descotes-Genon:2013wba,Buras:2013qja,Gauld:2013qja,Buras:2013dea,Altmannshofer:2014cfa,Celis:2015ara,Falkowski:2015zwa,Crivellin:2015mga,
Crivellin:2015lwa,Crivellin:2015era,Becirevic:2016zri,Boucenna:2016qad,Crivellin:2016ejn}), 
or by scalar or vector lepto-quarks (e.g.~\cite{Hiller:2014yaa,Gripaios:2014tna,Becirevic:2015asa,Varzielas:2015iva,Fajfer:2015ycq,Becirevic:2016yqi}). 
At one loop, they can be generated by box diagrams involving new particles (e.g.~\cite{Gripaios:2015gra,Bauer:2015knc,Arnan:2016cpy}) 
or by $Z^\prime$ penguins (e.g.~\cite{Belanger:2015nma}).
The step beyond the model-independent analysis allows to attempt a common explanation of the $b\to s\ell^+\ell^-$ anomalies with other tensions in flavour data,
like $R_{D^{(*)}}$ or the long-standing anomaly in the anomalous magnetic moment of the muon. It further permits to study the viability of the various
model classes in the light of constraints from other flavour observables and from direct searches. In the following, we will briefly summarize 
typical $Z^\prime$ and lepto-quark scenarios, and discuss bounds from $B_s-\overline{B}_s$ mixing and direct searches.

\boldmath
\subsection{$Z^\prime$ models}
\unboldmath
The interaction of a generic $Z^\prime$ boson with the SM fermions is described by the Lagrangian
\begin{equation}
 \mathcal{L}_{Z^\prime}\,=\,\sum\limits_{ff^\prime}\Gamma^L_{ff^\prime}\bar{f}\gamma^\mu P_L f^\prime Z^\prime_\mu\,+\,
                   \Gamma^R_{ff^\prime}\bar{f}\gamma^\mu P_R f^\prime Z^\prime_\mu\,+\,{\text h.c.},
                 \label{eq:ZLag}
\end{equation}
where the sum is over fermions $f,f^\prime$ with equal electric charges. The exact form
of the couplings $\Gamma^{L,R}_{ij}$ depends on the $U(1)^\prime$ charges assigned to the SM fermions
and on a potential embedding of the $Z^\prime$ in a more fundamental theory. Note, however, that
SU$(2)_L$ invariance implies the model-independent relations $\Gamma^L_{uu^\prime}=V_{ud}\Gamma^L_{dd^\prime}V^\dagger_{u^\prime d^\prime}$
and $\Gamma^L_{\ell\ell^\prime}=\Gamma^L_{\nu_\ell\nu_{\ell^\prime}}$ (with $V$ denoting the CKM matrix).

The Wilson coefficients $C_{9,10}^{(\prime)}$ are generated by tree-level $Z^\prime$ exchange involving products of couplings $\Gamma^{L,R}_{bq}\,\Gamma^{L,R}_{\ell\ell}$. Since only three out of these four products are independent, the relation $C_9\cdot C_{10}^\prime\,=\,C_9^\prime\cdot C_{10}$
is fulfilled in models with a single $Z^\prime$ boson. In order to generate a non-vanishing coupling $C_{9\,\mu}$, mandatory for a solution of the $b\to s\ell^+\ell^-$ anomalies, the couplings $\Gamma_{bs}^L$ and $\Gamma_{\mu\mu}^L+\Gamma_{\mu\mu}^R$ need to have non-vanishing values. 

The most popular class of $Z^\prime$ models is based on gauging $L_\tau-L_\mu$ lepton number~\cite{Altmannshofer:2014cfa,Crivellin:2015mga,Crivellin:2015lwa,
Altmannshofer:2015mqa,Altmannshofer:2016oaq}. This pattern of $U(1)^\prime$ charges avoids anomalies and is well-suited to generate
the measured PMNS matrix.  The vanishing coupling of the $Z^\prime$ to electrons allows to explain LFUV in
$R_K$ and helps to avoid LEP bounds on the $Z^\prime$ mass $M_{Z^\prime}$. The symmetry can be extended to the quark sector
 with a flavour non-universal assignment of $U(1)^\prime$ charges that induces the off-diagonal couplings $\Gamma^{L,R}_{bs}$ (
e.g.~\cite{Crivellin:2015lwa,Celis:2015ara}). An alternative mechanism to generate the couplings $\Gamma^{L,R}_{bs}$ consists in the introduction of additional vector-like quarks that are charged under the $U(1)^\prime$ symmetry 
and that generate an effective $bsZ^\prime$ coupling via their mixing with the SM fermions~\cite{Altmannshofer:2014cfa,Crivellin:2015mga,Altmannshofer:2015mqa,Bobeth:2016llm}. 

Several $Z^\prime$ scenarios have been proposed that are capable of solving not only the $b\to s\ell^+\ell^-$ anomalies but at the same time 
also other tensions in the data.
Embedding the $Z^\prime$ in a SU$(2)^\prime$ gauge model allows to address the anomalies in $R_{D^{(*)}}$ with a tree-level contribution to $b\to c\ell^-\bar{\nu}$ mediated by the $W^\prime$-boson~(e.g. \cite{Greljo:2015mma,Boucenna:2016qad}). It is also possible to solve the anomaly in $(g-2)_\mu$ in a $Z^\prime$ scenario, 
provided the $Z^\prime$ coupling to muons is generated at the loop-level so that both the NP contributions to $b\to s\ell^+\ell^-$ and 
$(g-2)_\mu$ are loop-suppressed~\cite{Belanger:2015nma}.

\boldmath
\subsection{Lepto-quark models}
\unboldmath
Lepto-quarks are new particles $\Delta_k$ beyond the SM that couple leptons to quarks via vertices $\ell_iq_j\Delta_k$. Different lepto-quark models can be classified according to the spin of the lepto-quarks and their quantum numbers with respect to the SM gauge groups. 
Since the $\ell_iq_j\Delta_k$ couplings violate lepton-flavour, lepto-quark models are excellent candidates to explain LFUV observables like $R_K$ 
and $R_{D^{(*)}}$. Indeed, various representations of lepto-quarks have been studied with respect to their capability of accommodating 
the measured values of $R_K$ and $R_{D^{(*)}}$ by tree-level lepto-quark contributions~\cite{Hiller:2014yaa,Gripaios:2014tna,Becirevic:2015asa,Varzielas:2015iva,Fajfer:2015ycq,Becirevic:2016yqi}. 
In Ref.~\cite{Bauer:2015knc} it was further proposed
that an SU$(2)_L$ singlet scalar lepto-quark could explain $R_{D^{(*)}}$ by a tree-level and $R_K$ by a loop contribution. This possibility was later
shown in Ref.~\cite{Becirevic:2016oho} to be challenged by other flavour data.

\boldmath
\subsection{Constraints from $B_s-\overline{B}_s$ mixing and direct searches}
\unboldmath
A NP model generating $b\to s\ell^+\ell^-$ necessarily also contributes to $B_{s}-\bar{B}_{s}$ mixing. 
In lepto-quark models, $b\to s\ell^+\ell^-$ is typically mediated at tree level, while $B_{s}-\bar{B}_{s}$ mixing contributions are loop-suppressed 
and thus do not pose relevant contraints. In $Z^\prime$ models, on the other hand, both processes are usually generated by 
tree-level exchange of the $Z^\prime$ boson. The constraint on $|\Gamma^{L,R}_{bs}|/M_{Z^\prime}$ from $B_{s}-\bar{B}_{s}$ mixing 
then imposes a lower limit of typically $|\Gamma^{L,R}_{\mu\mu}|/M_{Z^\prime}\gtrsim 0.3/(1\,\textrm{TeV})$ that needs to be reached for a solution of the 
$b\to s\mu^+\mu^-$ anomalies. In models with box contributions to both $b\to s\ell^+\ell^-$ and $B_s-\overline{B}_s$ mixing, the analogous constraint is more
severe due to the loop suppression: $|\Gamma_{\mu}|/\sqrt{M_{\Phi}}\gtrsim 3/\sqrt{1\,\textrm{TeV}}$ where $M_\Phi$ denotes the mass scale of the new particles in
the box and $\Gamma_\mu$ their coupling strength to the muon. It was shown in Ref.~\cite{Arnan:2016cpy} that this bound can be relaxed in a scenario 
with Majorana fermions in the box where the additional crossed boxes lead to a negative interference in $B_s-\overline{B}_s$ mixing.

Bounds from direct searches can be avoided to a large extent if the new physics couples only to the second and third fermion generation, in line with
LFUV in $R_K$ and $R_{D^{(*)}}$. Collider signals are then limited to more complex final state, like e.g. $pp\to 4\mu$ probing the muon-coupling of a
possible $Z^\prime$ boson, or to suppressed production channels, like e.g. $b\bar{s}\to\mu^+\mu^-$. However, it was found~\cite{Faroughy:2016osc} very recently that 
the data from Atlas/CMS already now heavily constrains $Z^\prime$ and lepto-quark scenarios even in the $b\bar{b}\to\tau^+\tau^-$ channel:
a solution of $R_{D^{(*)}}$ by SU$(2)^\prime$ gauge bosons $W^\prime/Z^\prime$ is restricted to masses $M_{Z^\prime}\lesssim 500\,\textrm{GeV}$,
and a solution via vector lepto-quarks is about to be excluded. The interplay with high-$p_T$ searches will thus definitely play a crucial role 
in the quest for an explanation of the flavour anomalies.

\section{Summary}
\label{sec:summary}

The discovery of the leptonic decay \decay{\Bs}{\mup\mun} by the CMS and LHCb collaborations was a major breakthrough of precision flavour physics with data from Run 1 of the Large Hadron Collider.
Since then, several anomalies have emerged in semileptonic decays.
These indicate a potential violation of lepton universality in the decays \decay{B}{K^{(*)}\mumu} and \decay{B}{K^{(*)}\ep\en} with a statistical significance of $2.6\sigma$.
In the angular analysis of \decay{B}{K^{*}\mumu} decays, tensions with the SM are seen in the longitudinal polarisation of the $K^*$ and the angular observable $P_5'$.
The largest discrepancy with more than $3\sigma$ is seen in the differential branching fraction of the decay \decay{\Bs}{\phi\mumu}.

The dynamics of \decay{B}{K^{*}\mumu} decays can be described by a set of helicity amplitudes, which, in the effective Hamiltonian formalism, split into Wilson coefficients and matrix elements of local operators.
Deviations in the Wilson coefficient $C_9$, which is sensitive to new physics as well as long-distance QCD, can explain the experimental data.
The most sensitive observables to $C_9$ are the $q^2$ dependence of the forward-backward asymmetry and the angular observable $P_5'$. 
These observables have been constructed such that they exhibit a reduced sensitivity to hadronic form factors, though a remnant dependence at order $\Lambda/m_{\bquark}$ cannot be avoided and its impact cannot be predicted in the heavy quark framework.
A combination of light-cone sum rule calculations with more precise measurements may be able to address this issue.
 
In light-cone sum rule computation finite width effects can effectively be 
bypassed if the vector mesons are treated consistently in all experiments, which includes 
those from where input is taken for form factor calculations as well as those where the 
form factor computations are used.  Whereas current lattice QCD studies do not include finite
width effects, recent developments indicate that this may change in the foreseeable future.
The use of equation of motion reduces the uncertainty of the projection 
on the \PB-meson state for ratios of form factors in light-cone sum rules. It is conceivable that the use of equation
of motion might help to further improve lattice QCD computation as well. 

The main focus of recent discussions have been the so-called charm contributions, which describe the sub-process $B \to  K^{(*)}(\bar c c \to \gamma^* \to \ell \ell)$.
The data are typically studied in different regions of $q^2$: the `partonic' well below the \jpsi resonance, the `narrow' between the \jpsi and $\psi(2S)$ resonances, and the `broad' in the high-$q^2$ region that is dominated by broad charmonium resonances.
Ideally, a coherent description of the low and high $q^2$ regions should be obtained.
Overcoming these challenges requires close collaboration of experimentalists and theorists to pursue new approaches such as a detailed study of the decays \decay{\PB}{\Dbar\PD K^{(*)}}.

Global fits aim to exploit a maximum of information of the range of observables in the framework of an effective theory.
This allows the splitting of the Wilson coefficients in a SM part and a component to encapsulate effects beyond the SM.
Most fits favour a non-SM value of the coefficient $C_9$ of about $-1$ with the pull of the SM scenario exceeding $4\sigma$ in several cases.
As contributions from high scales beyond the SM create $q^2$-independent effects, it is instructive to perform fits in several regions of $q^2$ and test for $q^2$-dependent effects that would indicate low-scale SM effects.
At the current level of precision these tests are consistent with a $q^2$-independent shift of $C_9$.

Possible explanations involving particles beyond the SM exist in the form of lepto-quarks or $Z^\prime$ bosons, typically mediating the \decay{\bquark}{\squark\ell^+\ell^-} transitions through tree-level exchange. 
In addition, these SM extensions have the potential to simultaneously accommodate other anomalies like $R_{D^{(*)}}$ or the anomalous magnetic moment of the muon. 
Direct searches pose tight constraints on some of these models and are expected to either probe or severely challenge them in the near future.

The number of flavour anomalies that appear to fit a common picture is intriguing.
The analysis of data taken during the ongoing Run 2 of the LHC will yield powerful new insight both into the observables of interest and into new strategies to control uncertainties.
The interpretation of these results requires close collaboration with theory, where advances are required in several areas for which promising strategies exist.

\bibliography{references,rareb,ffref}

\end{document}